\documentclass[%
11pt,
nofootinbib,
 amsmath,amssymb,
 aps,
prd,
]{revtex4-2}

\usepackage{graphicx}% Include figure files
\usepackage{dcolumn}% Align table columns on decimal point
\usepackage{bm}% bold math
\usepackage{array}
\usepackage{booktabs}
\usepackage{tabu}
\usepackage{dcolumn}
\usepackage{amsmath}
\usepackage{amsfonts}
\usepackage{amssymb}
\usepackage{graphicx}
\usepackage{subfigure}
\usepackage{graphicx}% Include figure files
\usepackage{dcolumn}% Align table columns on decimal point
\usepackage{bm}% bold math
\usepackage{xcolor}

\graphicspath{{figs/}} % Specify the path to the figures folder

\begin{document}

%\preprint{APS/123-QED}

\title{Non-commutativity in modified loop cosmology}% Force line breaks with \\

\author{Abolhassan Mohammadi}
    \email{abolhassanm@zjut.edu.cn; abolhassanm@gmail.com}
%\author{Second author}
    %\email{}
%\author{Third author}
    %\email{}
    
    \affiliation{Institute of Theoretical Physics and Cosmology, College of Science, Zhejiang University of Technology, Hangzhou, China.}



\date{\today}% It is always \today, today,

\begin{abstract}
In this study, we explore the pre-inflationary dynamics of the universe using a noncommutative extension of the mLQC-I framework. By incorporating a scalar field potential, we show that key features of Loop Quantum Cosmology (LQC), such as the quantum bounce and the super-inflationary phase, are preserved. Numerical solutions to the modified Hamiltonian equations, with initial conditions set at the quantum bounce, reveal that the universe's early expansion rate is sensitive to the shape of the potential. For a chaotic potential, the inclusion of noncommutativity results in a faster expansion rate, whereas for the Starobinsky potential, the expansion rate decreases with increasing noncommutative parameter $\theta$. Additionally, higher values of $\theta$ lead to an increased time derivative of the Hubble parameter, causing a shorter yet more expansive super-inflationary phase. Over time, Hubble parameters for different values of $\theta$ converge. For the Starobinsky potential, the Hubble parameter consistently decreases with larger $\theta$, resulting in a prolonged super-inflationary stage. The study also addresses the validation of the Hamiltonian constraint during the evolutionary time.
\end{abstract}

\maketitle

%\tableofcontents

%%%%%%%%%%%%%%%%%%%%%%%%%%%%%%%%%%%%%%%%%%%%%%
%%%%%%%%%%%%%%%%%%%%%%%%%%%%%%%%%%%%%%%%%%%%%%
%%%%%%%%%%%%%%%%%%%%%%%%%%%%%%%%%%%%%%%%%%%%%%
%%%%%%%%%%%%%%%%%%%%%%%%%%%%%%%%%%%%%%%%%%%%%%
%%%%%%%%%%%%%%%%%%%%%%%%%%%%%%%%%%%%%%%%%%%%%%
%%%%%%%%%%%%%%%%%%%%%%%%%%%%%%%%%%%%%%%%%%%%%%
\section{Introduction}
Without a doubt, Einstein's theory of gravity is the most successful framework we have for understanding gravitational phenomena, having passed numerous experimental and observational tests. However, the theory is incomplete, particularly in regimes of extremely high curvature, such as those present in the very early universe. It is widely accepted that a quantum theory of gravity is necessary to accurately describe the universe in these extreme conditions. Loop Quantum Gravity (LQG) stands as a leading candidate to provide a non-perturbative and background-independent quantization of gravity \cite{rovelli2004quantum,thiemann2008modern}. LQG introduces a quantum geometry where spatial areas and volumes are quantized, leading to a discrete spectrum of eigenvalues. Applying the techniques and methods of LQG to cosmological settings gives rise to Loop Quantum Cosmology (LQC) \cite{Singh:2009mz,Singh:2014fsy}.  \\ 

The concept of noncommutativity in spacetime was first introduced in 1947 by Snyder \cite{Snyder:1946qz} as a method to regularize quantum field theory. However, due to the success of the renormalization program, this concept was abandoned for an extended period. Noncommutative geometry was explicitly developed in the early 1980s \cite{Connes:1994yd}. TThe topic acquired considerable interest towards the end of the 20th century, partly due to breakthroughs in string theory \cite{Lee:1997uh, Seiberg:1999vs}. It was revealed that the low-energy limits of string theory naturally lead to noncommutative gauge theories. \cite{Connes:1997cr,Chu:1998qz,Schomerus:1999ug,Seiberg:1999vs,Garcia-Compean:2001jxk}. This discovery inspired widespread interest in the physical and mathematical ramifications of noncommutative field theory. \cite{Douglas:2001ba,Szabo:2001kg}. Furthermore, it is widely believed that at scales comparable to the Planck length, the continuum description of spacetime will break down in a complete theory of quantum gravity. One approach to model this breakdown is through the use of an uncertainty relation for spacetime coordinates, expressed as $\big[ x^i , x^j \big] = i \theta^{ij}$. Various noncommutative theories of gravity have been proposed, which incorporate this idea into the gravitational field \cite{Moffat:2000gr,Chamseddine:2000zu,Garcia-Compean:2003nix}. A noncommutative generalization of LQG was also constructed in \cite{Kober:2014wsa}. However, a common feature among these theories is their nonlinearity, which often makes them challenging to work with.  \\  

The two prominent candidates for quantum gravity, LQG and string theory, both suggest a departure from the conventional continuum description of spacetime. LQG, as a theory of quantum geometry, implies the discreteness of spacetime, while string theory, in its low-energy limit, leads to noncommutative gauge theories. This convergence of ideas has inspired efforts to construct a unified theory of quantum gravity that incorporates both discretization and noncommutativity. A straightforward approach to this, as explored in \cite{Espinoza-Garcia:2017qjl,Diaz-Barron:2019awc,Diaz-Barron:2021yha,Diaz-Barron:2023ctp,Diaz-Barron:2023qcf}, involves introducing a simplified form of noncommutativity within the framework of Loop Quantum Cosmology (LQC). Noncommutativity is thought to play a crucial role in the very early stages of the universe's evolution \cite{Diaz-Barron:2019awc}. Given this perspective, LQC provides an ideal framework for incorporating noncommutativity, as it naturally supports the generation of inflation and offers a clear understanding of pre-inflationary dynamics. Previous studies \cite{Espinoza-Garcia:2017qjl,Diaz-Barron:2019awc,Diaz-Barron:2021yha,Diaz-Barron:2023ctp,Diaz-Barron:2023qcf} have examined the effects of noncommutativity in LQC, both with and without a potential, on the early evolution of the universe. These works have demonstrated that key features of LQC, such as the quantum bounce, are preserved when noncommutativity is introduced. Moreover, noncommutativity has been shown to modify the dynamics, providing potential new insights into the early universe. In this paper, we take a step further by incorporating noncommutativity into the mLQC-I framework.  \\   

Because of the symmetry reduction before quantization in LQC, the Hamiltonian constraint differs from the one in LQG. Different Hamiltonians produce distinct Planck-scale physics. Attempts have been made based on Thiemann's proposal \cite{Thiemann:1996aw,Thiemann:1996av,Giesel:2006uj} to create Hamiltonians closer to the LQG structure. Yang et al. \cite{Yang:2009fp} suggested two effective Hamiltonians based on Thiemann's regularization. These modified Hamiltonians are commonly referred to as mLQC-I and mLQC-II.  \cite{Li:2018opr,Li:2018fco,Li:2019ipm,Li:2021mop,Li:2019qzr,Li:2020mfi}. Unlike standard LQC, where the Lorentzian term is proportional to the Euclidean term, mLQC-I treats these terms differently. In mLQC-I, the extrinsic curvature within the Lorentzian term is expressed in terms of holonomies, resulting in a fourth-order quantum difference equation, compared to the second-order constraint in standard LQC. Studies have explored cosmological dynamics, inflationary attractors, and other aspects of mLQC-I, revealing that due to the effective Hamiltonian and modified Friedmann equation, the universe's evolution is asymmetric about the bounce \cite{Li:2018opr,Li:2018fco,Li:2019ipm,Li:2019qzr,Li:2021mop,Li:2020mfi}. mLQC-I provides a more consistent framework for investigating quantum cosmology since it is more compatible with the fundamental concepts and methods of loop quantum gravity. Inspired by this consistency, we aim to extend previous work on noncommutativity in LQC \cite{Espinoza-Garcia:2017qjl,Diaz-Barron:2019awc,Diaz-Barron:2021yha,Diaz-Barron:2023ctp,Diaz-Barron:2023qcf} by investigating its effects within the mLQC-I framework. To investigate the behavior of the universe at very early times, two chaotic and Starobinsky potentials are considered, and it is found that the initial behavior of the universe around the bounce can be potential dependent, so that the noncommutativity with chaotic potential increases the universe expansion, while with Starobinsky, the universe experiences less expansion.   \\

This paper is organized as follows: in Sec. II, we briefly review the mLQC-I, introduce the noncommutativity, and then show how inclusion of the noncommutativity into the mLQC-I modifies the dynamical equations. The preinflationary dynamics of the universe will be considered in Sec. III for two potentials: chaotic and Starobinsky potentials. Due to the complexity of the dynamical equations, the analytical solutions are difficult to get; then, the numerical approaches are followed and the solutions are derived numerically. It is determined how inclusion of noncommutativity changes the behavior of the universe at very early times and affects the rate of expansion and also the number of e-folds. Finally, in Sec. IV, the result will be summarized. 

%%%%%%%%%%%%%%%%%%%%%%%%%%%%%%%%%%%%%%%%%%%%%%
%%%%%%%%%%%%%%%%%%%%%%%%%%%%%%%%%%%%%%%%%%%%%%
%%%%%%%%%%%%%%%%%%%%%%%%%%%%%%%%%%%%%%%%%%%%%%
%%%%%%%%%%%%%%%%%%%%%%%%%%%%%%%%%%%%%%%%%%%%%%
%%%%%%%%%%%%%%%%%%%%%%%%%%%%%%%%%%%%%%%%%%%%%%
%%%%%%%%%%%%%%%%%%%%%%%%%%%%%%%%%%%%%%%%%%%%%%
\section{mLQC-I and Non-commutativity}
In this section, we first briefly review the mLQC-I, its Hamiltonian, and the basic dynamical equations. Then, we introduce noncommutativity and determine how we are going to incorporate it into the theory. The noncommutative mLQC-I is presented in the last part, where the dynamical equations are modified due to the noncommutativity. In the next section, the model will be studied for two chaotic and Starobinsky potentials. 

%%%%%%%%%%%%%%%%%%%%%%%%%%%%%%%%%%%%%%%%%%%%%%
%%%%%%%%%%%%%%%%%%%%%%%%%%%%%%%%%%%%%%%%%%%%%%
%%%%%%%%%%%%%%%%%%%%%%%%%%%%%%%%%%%%%%%%%%%%%%
\subsection{Brief on mLQC-I}
The effective dynamics of the universe in loop cosmologies are determined via the effective Hamiltonian of the model. The effective Hamiltonian is a combination of a Euclidean term and a Lorentzian term. In the standard LQC, the Lorentzian term is assumed to be proportional to the Euclidean term, so that the effective Hamiltonian is given as 
\begin{equation}
	\mathcal{H} = -\frac{3v}{8 \pi G \lambda^2 \gamma^2} \; 
	\sin^2\left(\lambda \beta \right) + \mathcal{H}_{\phi}.
\end{equation} 
where $\mathcal{H}_\phi$ is the Hamiltonian of the scalar field, read as $\mathcal{H}_\phi = p_\phi^2 / 2 v + v \; V(\phi)$, $p_\phi$ is the momentum conjugate of the scalar field, and $V(\phi)$ is the potential of the field. \\ 
In modified LQCs, there is no such relation between the Lorentzian and the Euclidean parts of the Hamiltonian, and they are treated differently by following the actual construction of LQG. In mLQC-I, the extrinsic curvature in the Lorentzian term is expressed in terms of the connection and the volume, so that
\begin{equation}
	K^i_a=\frac{1}{\kappa \gamma^3}\{A^i_a,\{\mathcal{H}_{\mathrm{grav}}^{(E)},V\}\},
\end{equation}  
Substituting this expression back into the gravitational part of the Hamiltonian, the final effective Hamiltonian of the theory is achieved as
\begin{equation}
	\mathcal {H}_{\mathrm{I}} =\frac{3v}{8\pi G\lambda^2}\left\{\sin^2(\lambda \beta)-\frac{(\gamma^2+1)\sin^2(2\lambda \beta)}{4\gamma^2}\right\} + \mathcal{H}_\phi,
\end{equation}
The dynamical equations are acquired from the above effective Hamiltonian
\begin{eqnarray}
\dot v &=& \Big\{v, \mathcal H_{\mathrm{I}}\Big\}=\frac{3v\sin(2\lambda \beta)}{2\gamma \lambda}\Big\{(\gamma^2+1)\cos(2\lambda \beta)-\gamma^2\Big\}, \\ 
\dot \beta &=& \Big\{b, \mathcal H_{\mathrm{I}}\Big\}=\frac{3\sin^2(\lambda \beta)}{2\gamma \lambda^2}\Big\{\gamma^2\sin^2(\lambda \beta)-\cos^2(\lambda \beta)\Big\} - 4\pi G\gamma P_\phi, \\
\dot\phi &=& \Big\{\phi, \mathcal H_{\mathrm{I}}\Big\} = \frac{p_{\phi}}{v}, \\
\dot{p}_\phi &=& \Big\{p_\phi, \mathcal H_{\mathrm{I}}\Big\} = -v \; V'(\phi),
\end{eqnarray}
where prime in the last equation stands for derivative with respect to the scalar field, and $P_\phi$ is the pressure of the scalar field defined, $P_\phi = - \partial{\mathcal{H}_I} / \partial v = \dot{\phi}^2 / 2 v^2 + V(\phi)$.  From the Hamiltonian constraint, one can get the energy density as follow:
\begin{equation}\label{rho_mlqc1}
	\rho=\frac{3}{8\pi G\lambda^2}\left(-\sin^2(\lambda \beta)+\frac{(\gamma^2+1)\sin^2(2\lambda \beta)}{4\gamma^2}\right), 
\end{equation}
leading us to
\begin{equation}
	\sin^2(\lambda \beta_{\pm})= \frac{1\pm\sqrt{1-\rho/\rho_c^I}}{2(\gamma^2+1)},
\end{equation} 
where $\rho_c^I$ is the critical energy density given as $\rho_c^I \equiv3/[32\pi \lambda^2\gamma^2(\gamma^2+1)G]$. From this equation, it is concluded that there are two possible solutions for the parameter $\beta$, so that for the contracting phase only $\beta_{-}$ is acceptable and for the expanding universe only the solution $\beta_{+}$ is the valid solution \cite{Li:2018opr}. This affects the Friedmann equation and indicates that the Friedmann equations for the contracting and expanding universes are not the same \cite{Li:2018opr}. In other words, there is asymmetry in the evolution of the universe around the bounce.

%%%%%%%%%%%%%%%%%%%%%%%%%%%%%%%%%%%%%%%%%%%%%%
%%%%%%%%%%%%%%%%%%%%%%%%%%%%%%%%%%%%%%%%%%%%%%
%%%%%%%%%%%%%%%%%%%%%%%%%%%%%%%%%%%%%%%%%%%%%%
\subsection{Noncommutativity}
Quantum cosmology is an attempt to determine the quantum behavior of the universe at the very early stages of its evolution. The Wheeler-DeWitt (WDW) equation presents the quantum behavior of the universe in the standard quantum cosmology \cite{Bojowald:2020nwa,Bojowald:2011zzb,Calcagni:2017sdq,Jalalzadeh:2020bqu}. Due to the existence of symmetries in these models, the degrees of freedom of the gravitational field and matter field are reduced to a finite number, leading to a finite-dimensional phase space known as minisuperspace \cite{ryder:1972,Hartle:1983ai}. Following this procedure, one can get a simple and approximate model of quantum gravity \cite{Garcia-Compean:2001jxk}. Another approach for considering the quantum mechanics effects in the early universe evolution is the idea of introducing deformations to the Poisson bracket \cite{Bayen:1977pr,Bayen:1977ha,Bayen:1977hb} as a part of a consistent and complete type of quantization known as deformation quantization. The deformed minisuperspace, as an attempt to incorporate effective noncommutativity, was first considered in \cite{Garcia-Compean:2001jxk} with the name of noncommutative cosmology. The deformation quantization is also imposed on the flat minisuperspace in \cite{Cordero:2011xa}. Hence, analyzing cosmological models in a deformed phase space can be viewed as an exploration of quantum effects on cosmological solutions \cite{Vakili:2010qf,Malekolkalami:2014dca,Rashki:2014noa,Jalalzadeh:2017jdo,Bina:2010ir}. \\ 
The Moyal brackets $\{f,g\}_{\alpha}=f\star_{\alpha}g-g\star_{\alpha}f$ encodes the phase space deformation. Here, the Moyal product \cite{Gamboa:2000yq,Chaichian:2000si}
\begin{equation}\label{Moyal_product}
    (f\star g)(x)=\exp\left[\frac{1}{2}\alpha^{ab}\partial_a^{(1)}\partial_b^{(2)}\right]f(x_1)g(x_2)\bigg|_{x_1=x_2=x},
\end{equation}
is introduced instead of the usual product of functions, in which, $\alpha$ is given by
\begin{eqnarray}\label{Moyal_product_alpha}
\alpha =
\left( {\begin{array}{cc}
 \theta_{ij} & \delta_{ij}+\sigma_{ij}  \\
- \delta_{ij}-\sigma_{ij} & \eta_{ij}  \\
 \end{array} } \right),
\end{eqnarray}
where $\theta_{ij}$ and $\eta_{ij}$ are two $2 \times 2$ antisymmetric tensors that respectively describe the noncommutativity between coordinates and momenta. The matrix $\sigma_{ij}$ is also defined as $\sigma_{ij}=-\frac{1}{8}(\theta_{i}^k\beta_{kj}+\beta_{i}^k\theta_{kj})$. The algebra related to this phase space variables is known as the $\alpha$-deformed algebra, given by
\begin{equation}\label{alpha_deformed_algebra}
\{x_i,x_j\}_{\alpha}=\theta_{ij}, \quad \{x_i,p_j\}_{\alpha}=\delta_{ij}+\sigma_{ij},\quad \{p_i,p_j\}_{\alpha}=\eta_{ij},
\end{equation}
There is another way to make the similar algebra as above. Applying the following transformation on the variables of the phase space
\begin{eqnarray}\label{phase_space_variabel}
\hat{x}=x+\frac{\theta}{2}p_{y}, \qquad \hat{y}=y-\frac{\theta}{2}p_{x}, \nonumber \\
\hat{p}_{x}=p_{x}-\frac{\theta}{2}y, \qquad \hat{p}_{y}=p_{y}+\frac{\theta}{2}x,
\end{eqnarray}
One can confirm that the phase space transformed variables satisfy the following deformed algebra:
\begin{equation}\label{new_algebra}
	\{\hat{y},\hat{x}\}=\theta,\quad 
	\{\hat{x},\hat{p}_{x}\}=\{\hat{y},
	\hat{p}_{y}\}=1+\sigma,\quad 
	\{\hat{p}_y,\hat{p}_x\}=\theta,
\end{equation}
in which we have used the expressions $\theta_{ij}=-\theta\epsilon_{ij}$, $\beta_{ij}=\beta\epsilon_{ij}$, and $\sigma=\theta\eta/4$, where $\epsilon_{ij}$ is the Levi-Civita symbol. It is evident that these relations mirror those in Eq.\eqref{alpha_deformed_algebra}; however, the brackets in Eq.\eqref{new_algebra} are considered to be the conventional Poisson brackets, whereas those in Eq.\eqref{alpha_deformed_algebra} are known as the $\alpha$-deformed Poisson brackets. \\
The canonical Hamiltonian $\mathcal{H}$ in the deformation quantization formalism, where the $\alpha$-deformed algebra is imposed on the system, changes to the deformed Hamiltonian $\mathcal{H}'$. On the other hand, in the deformed phase space described by Eq,\eqref{new_algebra}, it is always possible to construct a Hamiltonian analogous to the canonical one. Practically, it is easier and more convenient to work with the latter approaches. Here, the Hamiltonian keeps the same functional form so that the phase space variables $(x, y, p_x, p_y)$ are replaced by the transformed ones $(\hat{x}, \hat{y}, \hat{p}_x, \hat{p}_y)$. Then, the equations of motion are written for the original phase space variables $(x, y, p_x, p_y)$, that include modification terms related to the noncommutative effects.

From two approaches for formulating the deformed algebra that can be found in the literature, i.e., applying the $\alpha$-algebra to the system and introducing the shifted variables, the latter has been followed in different cosmological models \cite{Perez-Payan:2011cvf,Perez-Payan:2014kea,Lopez:2017xaz,Vakili:2010qf,Malekolkalami:2014dca}. The advantage of the method is that a deformation of the Poisson bracket is not required. The noncomutative Hamiltonian, $\mathcal{H}^{nc}$, also has the same functional form as the Hamiltonian $\mathcal{H}$, but with a rearrangement of the variables so that the algebra \eqref{new_algebra} be satisfied. A deformed algebra, same as the ones in Eq.\eqref{new_algebra}, in the phase space spanned by the main variables $\{ v, b, \phi, \pi_\phi \}$, only on the momentum sector as
\begin{equation}\label{nc_poisson}
\{\beta^{nc},v^{nc}\}=4\pi G\gamma,\quad 
\{v^{nc},p^{nc}_{\phi}\}=\theta, \quad 
\{\phi^{nc},p^{nc}_{\phi}\}=1,
\end{equation}
will be the start of our work where we have only consider the noncommutativity in the momentum sector. These relations in the model can also be applied by utilizing the following shifted variables:
\begin{equation}\label{shifted_variables}
	\beta^{nc}=\beta, \quad 
	\phi^{nc}=\phi, \quad 
	v^{nc}=v + a \theta \phi, \quad 
	p_{\phi}^{nc}=p_\phi + b \theta \beta,
\end{equation}
where $a$ and $b$ are two constants that follow the relation $a-4\pi G\gamma b=1$ \cite{Espinoza-Garcia:2017qjl,Diaz-Barron:2019awc}.  \\
The effective Hamiltonian is the basic equation for our model, so the dynamical equations are constructed based on the effective Hamiltonian. Therefore, to build our deformed theory, we begin with the effective Hamiltonian. In the following subsections, we introduce the deformed effective Hamiltonian in mLQC-I, and then the main dynamical equations of the models are introduced. \\

\subsection{Noncomutativity in mLQC-I}
Including the noncomutativity to the model, the effective deformed Hamiltonian has the same functional shape 
\begin{equation}
	\mathcal{H}_{\mathrm{I}}^{nc} =
  \frac{3v^{nc}}{8\pi G\lambda^2} \left\{ \sin^2(\lambda \beta) - 
       \frac{(\gamma^2+1) \sin^2(2\lambda \beta)}{4\gamma^2}  \right\} 
            + \frac{\big( p_\phi^{nc} \big)^2}{2 v^{nc}} + v^{nc} \; V(\phi).
\end{equation}
Following the same process, the dynamical equations are extracted from the above Hamiltonian so that
\begin{eqnarray}
	\dot{v}  & = &  \frac{3v\sin(2\lambda \beta)}{2\gamma \lambda}\Big\{(\gamma^2+1)\cos(2\lambda \beta)-\gamma^2\Big\} - \big( 4\pi G \gamma b \theta \big) \; \frac{p_\phi^{nc}}{v^{nc}}, \label{v1_equation}   \\
	\dot{\beta}  & = & \frac{3\sin^2(\lambda \beta)}{2\gamma \lambda^2}\Big\{\gamma^2\sin^2(\lambda \beta)-\cos^2(\lambda \beta)\Big\} - 4\pi G\gamma P,  \label{b1_equation}   \\
	\dot{\phi}  & = & \frac{p_\phi + b \theta \beta}{v + a \theta \phi},  \label{phi1_equation} \\
	\dot{p}_\phi & = & - \frac{3 a \theta}{8 \pi G \gamma^2 \lambda^2} \; \left\{ \sin^2(\lambda \beta) - 
       \frac{(\gamma^2+1) \sin^2(2\lambda \beta)}{4\gamma^2}  \right\} - a \theta \frac{(p^{nc}_\phi)^2}{2(v^{nc})^2} - a \theta \; V(\phi) - v^{nc} \; V'(\phi) \label{pi1_equation} 
\end{eqnarray}
Taking $\theta \rightarrow 0$, the equations reduce to ones in the commutative mLQC-I. The energy density is obtained from the Hamiltonian constraint and Eq.\eqref{pi1_equation}, which is found to have the same functional form in terms of $\beta$ as Eq.\eqref{rho_mlqc1}, so that
\begin{equation}\label{rho_nc_mlqc1}
	\rho= \frac{(p_\phi^{nc})^2}{2 (v^{nc})^2} + V(\phi) = \frac{3}{8\pi G \gamma^2 \lambda^2}\left(-\sin^2(\lambda \beta)+\frac{(\gamma^2+1)\sin^2(2\lambda \beta)}{4\gamma^2}\right).  
\end{equation}
Reversing the equation, it is found out that there are the same two acceptable values for $\beta$, indicating that the evolution equation are different for the contracting and expanding phases of the universe. The critical energy density at the bounce also remains the same. \\ 

Eq.\eqref{rho_nc_mlqc1} indicates that the energy density has a maximum as $\rho_c^I$; however, at what time the energy density reaches this value might be different from the commutative mLQC-I. This is due to the presence of the extra term $\big( 4\pi G \gamma b \theta \big) \; p_\phi^{nc} / v^{nc}$ in Eq.\eqref{v1_equation}. In general, we expect the same behavior so that the energy density reaches the maximum value $\rho_c^I$ at the bounce, where $\dot{v} = 0$, But there might be a shift in the time of bounce occurrence. Therefore, in general, it could be concluded that although the functional form of the above energy density in terms of the cosmic time $t$ in general may be different due to the presence of the noncommutative parameter $\theta$ that may vary the behavior of $\beta(t)$. The Eq.\eqref{b1_equation} is the same as the one in commutative mLQC-I; however, the other equations are getting more complicated and different functional form in terms of time $t$ is expected.   \\ 
There are two subsets of Eq.\eqref{shifted_variables}, which are mainly considered in the literature \cite{Espinoza-Garcia:2017qjl,Diaz-Barron:2019awc,Diaz-Barron:2021yha,Diaz-Barron:2023ctp,Diaz-Barron:2023qcf}. The first choice is the simultaneous fulfillment of 
\begin{equation}
	b = 0, \qquad v^{nc}(t_B) > 0,
\end{equation}
where $t_B$ stands for the time when the energy density reaches the maximum energy density $\rho_c^I$. Besides, the above is also a sufficient condition to have a minimum volume at the time when the energy density is equal to $\rho_c^I$; it can be checked from Eq.\eqref{v1_equation}. The second subset is the fulfillment of the following condition:
\begin{equation}
	\dot{\phi}(t_B) = 0, \qquad b \theta \ddot{\phi} < 0,
\end{equation}
that is also a sufficient condition to have a minimum volume bounce, where the energy density reaches its maximum value. This condition allows us to fix the value of the scalar field at the bounce because, at this point, the energy density is equal to the potential part, and $\rho(t_B) = V(\phi(t_B)) = \rho_c^I$. On the other hand, the momentum conjugate of the scalar field at the bounce is obtained using Eq.\eqref{phi1_equation}, in which $p_\phi(t_B) = - b \theta \beta(t_B)$ and $\sin(\lambda \beta(t_B)) = 1 / 2(\gamma^2 + 1)$. \\
These two conditions might not cover all possible cases of having the minimum bounce, but they definitely cover all possible cases of having a minimum volume bounce at $\rho = \rho_c$. In the next section, we are going to consider the solution by including the above two conditions. 

\section{Numerical Solutions}
The main dynamical equations of the models, introduced in the last section, are complicated, and getting an analytical solution is difficult and complex. Therefore, we go for the numerical result and determine the behavior of the main quantities of the universe's evolution. The investigation is confined to two subsets of Eq.\eqref{shifted_variables} introduced in the previous section. In both cases, we have the maximum of the energy density at the bounce point where the universe has a minimum volume, i.e., $\dot{v} = 0$. \\

%%%%%%%%%%%%%%%%%%%%%%%%%%%%%%%%%%%%%%%%%%%%%%
%%%%%%%%%%%%%%%%%%%%%%%%%%%%%%%%%%%%%%%%%%%%%%
%%%%%%%%%%%%%%%%%%%%%%%%%%%%%%%%%%%%%%%%%%%%%%
%%%%%%%%%%%%%%%%%%%%%%%%%%%%%%%%%%%%%%%%%%%%%%
\subsection{Condition I: Chaotic potential}
\begin{figure}
    \centering
    \subfigure[]{\includegraphics[width=7cm]{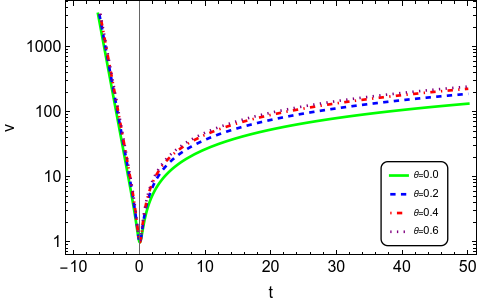}}
    \subfigure[]{\includegraphics[width=7cm]{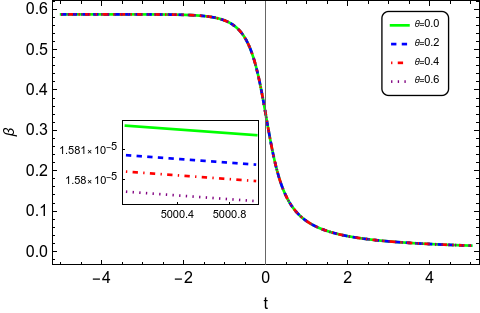}}
    \subfigure[]{\includegraphics[width=7cm]{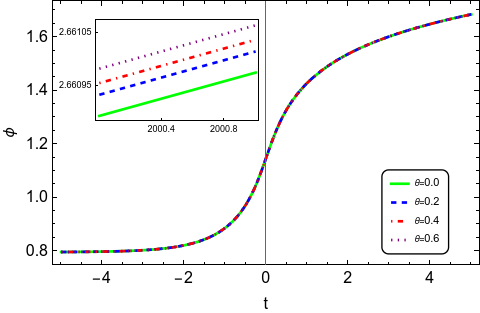}}
    \subfigure[]{\includegraphics[width=7cm]{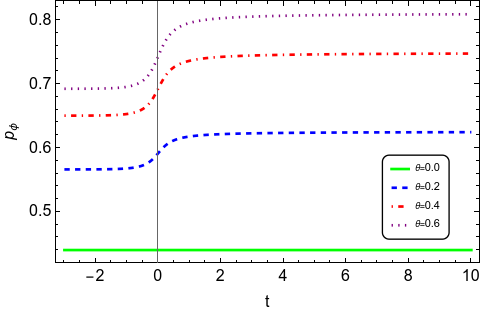}}
    \caption{The plot shows the behavior of the a) volume b) function $\beta$, c) scalar field, and d) the field momentum conjugate versus time for different values of the noncommutative parameter $\theta$. With an increase in the parameter $\theta$, the volume of field momentum conjugate enhance as well. Also, $\theta$ slightly increases $\beta$ and $\phi$.}
    \label{c1_main_quanities}
\end{figure}

\begin{figure}
    \centering
    \subfigure[]{\includegraphics[width=7cm]{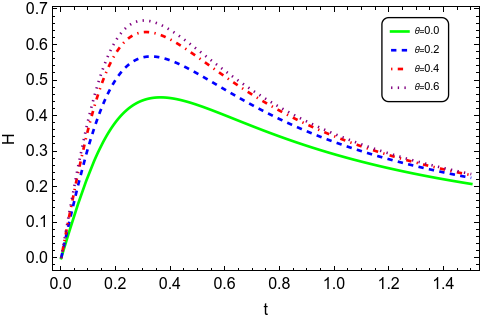}}
    \subfigure[]{\includegraphics[width=7cm]{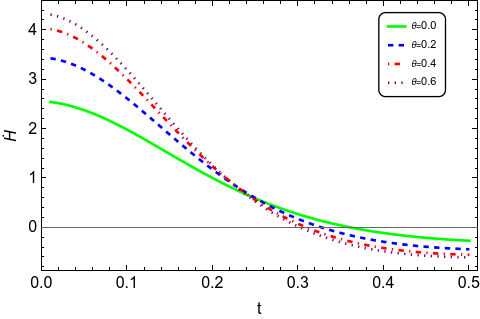}}
    \caption{The behavior of the Hubble parameter and its time derivative $\dot{H}$ is plotted versus time for different values of $\theta$. Increasing $\theta$ leads to a bigger $\dot{H}$ that result in a larger Hubble parameter with higher peak. Also, for bigger value of $\theta$, $\dot{H}$ happens sooner.  }
    \label{c1_HdH}
\end{figure}

\begin{figure}
    \centering
    \includegraphics[width=7cm]{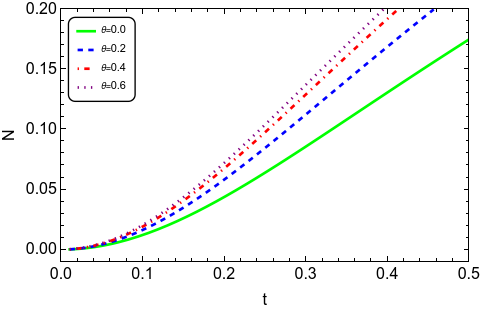}
    \caption{The number of e-fold of the universe in the first initial time after the bounce versus time for different values of $\theta$ is illustrated. It is found that bigger $\theta$ gives more number of e-fold.}
    \label{c1_efold_superinflation}
\end{figure}

\begin{figure}
    \centering
    \includegraphics[width=7cm]{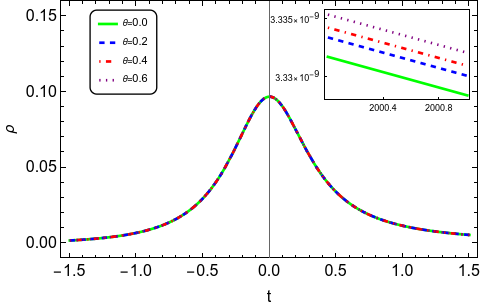}
    \caption{The energy density of the scalar field versus time for different values of $\theta$. The parameter $\theta$ has a slight effect on the magnitude of the energy density, however, the general behavior remains the same.}
    \label{c1_energy_density}
\end{figure}

\begin{figure}
    \centering
    \includegraphics[width=7cm]{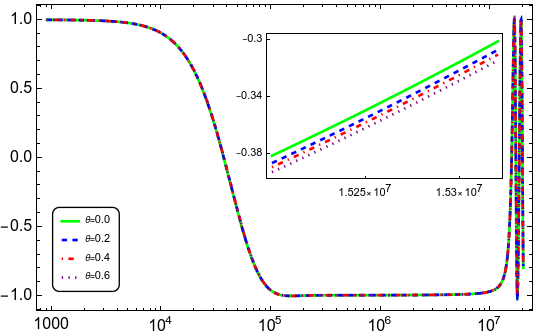}
    \caption{The equation of state parameter $\omega = \rho / P$ versus time for different values of $\theta$. The general behavior of the parameter is the same for different values of the parameter $\theta$, while one could see a small difference which indicates that the inflationary phase ends sooner for smaller valuer of $\theta$. }
    \label{c1_omega}
\end{figure}

\begin{figure}
    \centering
    \includegraphics[width=7cm]{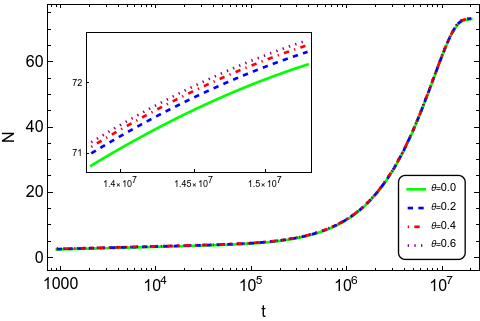}
    \caption{The figure shows the total number of e-folds for different values of $\theta$. One can see that the universe expands more for bigger values $\theta$, however, the difference is not tangible.}
    \label{c1_efold_total}
\end{figure}

\begin{figure}
    \centering
    \includegraphics[width=8cm]{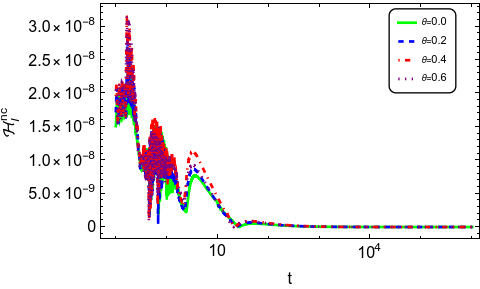}
    \caption{The figure shows the Hamiltonian of the model versus the cosmic time $t$ for different values of $\theta$. As it can be seen, the Hamiltonian for all time with a good approximation is close to zero. This confirms the validity of the Hamiltonian constraint with good approximation. }
    \label{c1_Hamiltonian}
\end{figure}

This condition is related to the subset of Eq.\eqref{shifted_variables} where $b = 0$ and $v(t_B) > 0$. This condition ensures that at the time $t_B$ where the energy density reaches the maximum $\rho_c^I$, the volume of the universe is at its minimum. Using this, we are going to set the initial condition at the bounce time, following the same procedure as the commutative mLQC-I. Due to the scale symmetry of the model and without loss of generality, the volume of the universe at the bounce point is set to be $v(t_B) = v_B = 1$. Choosing the field at the bounce time to be $\phi(t_B) = \phi_B$, the field momentum conjugate $p_\phi(t_B)$ could be read from the energy density, where $\rho(t_B) = \big( p^{nc}_\phi(t_B) \big)^2 / 2 \big( v^{nc}(t_B) \big)^2 + V(\phi_B) = \rho_c^I$. From the Hamiltonian constraint, the function $\beta$ at the bounce is obtained to be $\beta(t_B) = \beta_B = \arcsin( 1 / \sqrt{2 (\gamma^2 + 1)} / \lambda$. Applying these initial conditions to the differential equations \eqref{v1_equation}, \eqref{b1_equation}, \eqref{phi1_equation}, and \eqref{pi1_equation}, the main quantities of the model are evaluated. Fig.\ref{c1_main_quanities} shows the volume, $\beta(t)$, the scalar field $\phi(t)$, and its momentum conjugate $p_\phi(t)$ versus time for different values of the noncomuutative parameter $\theta$. At first glance, it is seen that the scalar field $\phi(t)$ and the function $\beta(t)$ are not affected by the commutative parameter $\theta$; however, by extending the time and focusing on the lines, a small difference shows up, implying a smaller (bigger) value of $\beta$ ($\phi$). On the other side, the parameter $\theta$ changes the behavior of the universe volume $v(t)$ and the scalar field momentum conjugate $p_\phi(t)$. The figure shows that the volume of the universe is higher for larger values of $\theta$, and also $p_\phi$ gets a larger value with an enhancement of $\theta$. \\ 
%%%%%%%%%%%%%%%%%%%%%%%%%%%%%%%%%
%%%%%%%%%%%%%%%%%%%%%%%%%%%%%%%%%
The Hubble parameter and its time derivative are illustrated in Fig.\ref{c1_HdH} for different values of $\theta$. There is a stage of super-inflation where the $\dot{H}$ is positive and the Hubble parameter increases. By increasing the noncommutative parameter $\theta$, $\dot{H}$ gets bigger, indicating a faster enhancement for the Hubble parameter, which is clearly exhibited in the figure. It can also be realized that $\dot{H}$ related to bigger $\theta$ decreases faster and the super-inflationary stage ends earlier by increasing $\theta$. 
%%%%%%%%%%%%%%%%%%%%%%%%%%%%%%%%%
%%%%%%%%%%%%%%%%%%%%%%%%%%%%%%%%% 
The expansion of the universe during the super inflationary stage is plotted in Fig.\ref{c1_efold_superinflation}, where it is found that the universe expands a little more by taking a bigger value of $\theta$ (remember that the super-inflationary stage occurs in a very short time right after the bounce). \\
%%%%%%%%%%%%%%%%%%%%%%%%%%%%%%%%% 
%%%%%%%%%%%%%%%%%%%%%%%%%%%%%%%%% 
The noncommutative parameter $\theta$ has small affection on the scalar field, mainly at a later time. Therefore, a small difference in the energy density of the field corresponding to different values $\theta$ is expected not for the initial time after the bounce but rather for the later time. Fig.\ref{c1_energy_density} exhibits the behavior of the energy density, and it is seen that all plots overlap at the initial time; however, by passing time, a small difference between the curves is detected. 
%%%%%%%%%%%%%%%%%%%%%%%%%%%%%%%%% 
%%%%%%%%%%%%%%%%%%%%%%%%%%%%%%%%% 
The equation of the state parameter depends on the scalar field and its time derivative. Since by varying $\theta$ there is a small change in the field, the equation of the state parameter is expected to show only a small change for each selected value of $\theta$, which is plotted in the inset plot of Fig.\ref{c1_omega}. The figure states that if we take $\omega = -1/3$ as the end point of inflation\footnote{There are different ideas on the definition of the start time and end point of the inflationary phase in literature, e.g., see \cite{Baumann:2009ds,Agullo:2015tca,Shahalam:2017wba,Li:2019ipm,Xiao:2020olb}. Following \cite{Zhu:2017jew}, the end point of the inflationary phase is taken as $\omega = -1/3$.}, the inflationary phase will end sooner for a larger value of $\theta$. 
%%%%%%%%%%%%%%%%%%%%%%%%%%%%%%%%% 
%%%%%%%%%%%%%%%%%%%%%%%%%%%%%%%%%  
The total number of e-folds is shown in Fig.\ref{c1_efold_total}, where the difference in $N$ for different values of $\theta$ is emphasized in the inset plot. The end time of the inflationary phase for a higher value of $\theta$ is bigger, which results in a bigger e-fold of expansion. However, reading the end time of inflation as $t = 1.55 \times 10^7$ from Fig.\ref{c1_omega}, it is found that the differences in the total number of e-fold for different values of $\theta$ are not tangible. This value is changing between $72$ and $73$.  \\ 
%%%%%%%%%%%%%%%%%%%%%%%%%%%%%%%%% 
%%%%%%%%%%%%%%%%%%%%%%%%%%%%%%%%%  
The Hamiltonian constraint in (m)LQC, where the main dynamics of the theory is derived from the effective Hamiltonian, is an essential feature stating the Hamiltonian must be equal to zero, $\mathcal{H}^{nc} = 0$. This constraint is a fundamental aspect of the theory, ensuring the consistency of the dynamical evolution. Fig.\ref{c1_Hamiltonian} displays the Hamiltonian versus the cosmic time. The Hamiltonian exhibited in the figure is not exactly zero, and it is at most of the order of $10^{-8}$. It should be mentioned that since here we followed the numerical method, small deviation from the zero is expected and this deviation is in nature of the numerical approaches. The slightly numerical discrepancy we see in the figure is acceptable and it safely can be concluded that the Hamiltonian constraint is satisfied. 
%%%%%%%%%%%%%%%%%%%%%%%%%%%%%%%%% 
%%%%%%%%%%%%%%%%%%%%%%%%%%%%%%%%%  

%%%%%%%%%%%%%%%%%%%%%%%%%%%%%%%%%%%%%%%%%%%%%%
%%%%%%%%%%%%%%%%%%%%%%%%%%%%%%%%%%%%%%%%%%%%%%
%%%%%%%%%%%%%%%%%%%%%%%%%%%%%%%%%%%%%%%%%%%%%%
%%%%%%%%%%%%%%%%%%%%%%%%%%%%%%%%%%%%%%%%%%%%%%
\subsection{Condition I: Starobinsky potential}
\begin{figure}
    \centering
    \subfigure[]{\includegraphics[width=7cm]{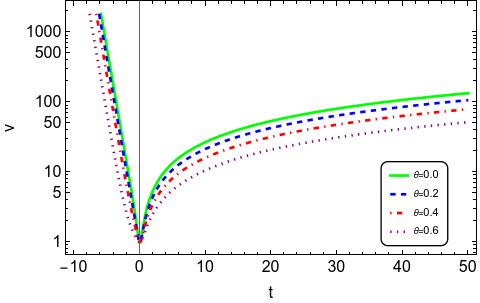}}
    \subfigure[]{\includegraphics[width=7cm]{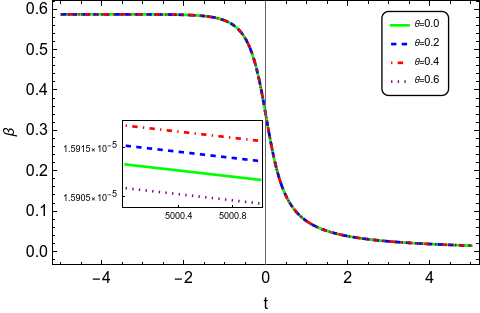}}
    \subfigure[]{\includegraphics[width=7cm]{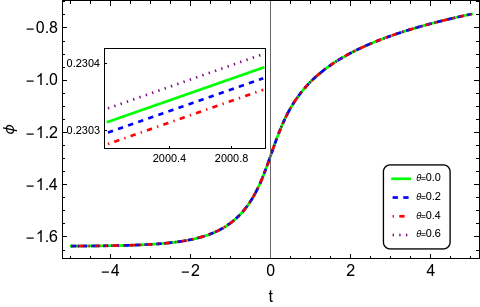}}
    \subfigure[]{\includegraphics[width=7cm]{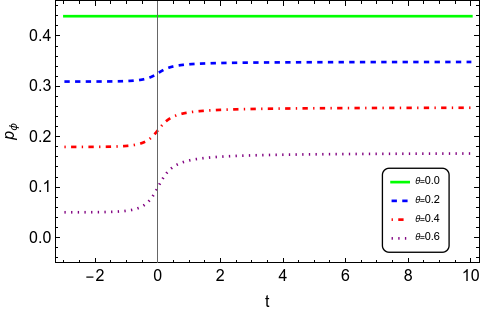}}
    \caption{The plot shows the behavior of the a) volume b) function $\beta$, c) scalar field, and d) the field momentum conjugate versus time for different values of the noncommutative parameter $\theta$ where the potential is chosen to be Starobinsky potential. Unlike the previous case, the volume and the field momentum conjugate decrease by enhancement of $\theta$. The scalar field and the funciton $\beta$ experience slightly difference by variation of $\theta$.  }
    \label{c1Strb_main_quanities}
\end{figure}

\begin{figure}
    \centering
    \subfigure[]{\includegraphics[width=7cm]{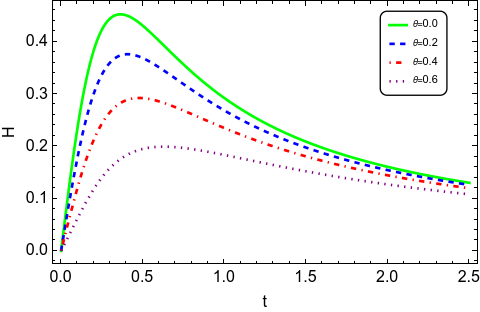}}
    \subfigure[]{\includegraphics[width=7cm]{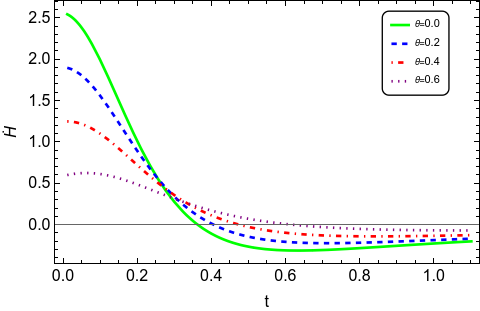}}
    \caption{The behavior of the Hubble parameter and its time derivative $\dot{H}$ is plotted versus time for different values of $\theta$. Smaller $\dot{H}$ and Hubble parameter is given with an increase in $\theta$. The decreasing rate enhances by reducing $\theta$ which gives shorter super-inflationary phase for bigger values of $\theta$. }
    \label{c1Strb_HdH}
\end{figure}

\begin{figure}
    \centering
    \includegraphics[width=7cm]{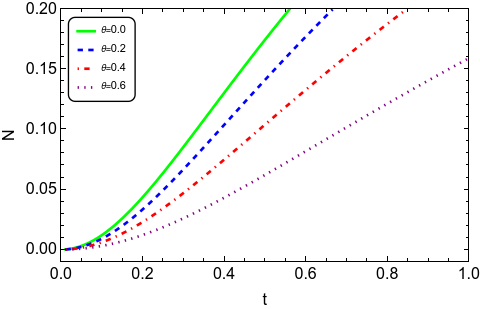}
    \caption{The number of e-fold of the universe for the initial times after the bounce is given for different values of $\theta$. Higher expansion is provided by smaller value of $\theta$. }
    \label{c1Strb_efold_superinflation}
\end{figure}

\begin{figure}
    \centering
    \includegraphics[width=7cm]{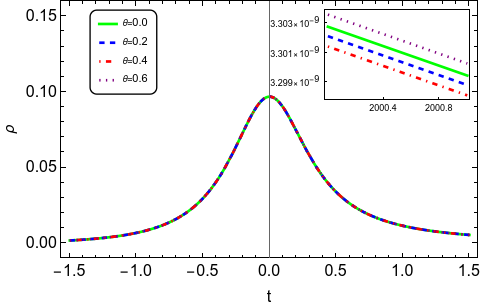}
    \caption{The energy density of the scalar field versus time for different values of $\theta$ which shows the same behavior. There is a slight difference in the magnitude of the energy density by varying $\theta$.}
    \label{c1Strb_energy_density}
\end{figure}

\begin{figure}
    \centering
    \subfigure[\label{c1Strb_omega}]{\includegraphics[width=7cm]{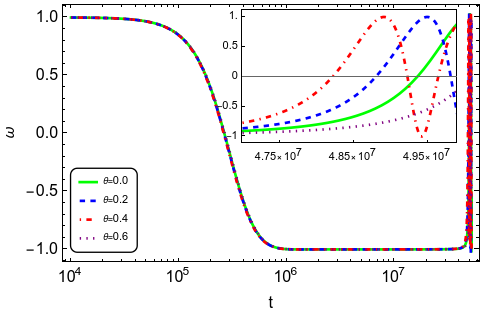}}
    \subfigure[\label{c1Strb_efold_total}]{\includegraphics[width=7cm]{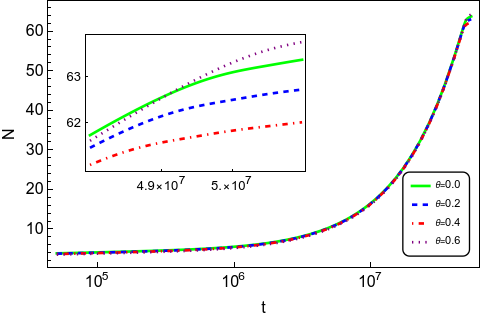}}
    \caption{a)The equation of state parameter $\omega = \rho / P$ and b) the total number of e-folds versus time is plotted for different values of $\theta$. The general behavior of the parameters are the same, however, one can see a small differences by changing $\theta$ so that the end time of inflation changes that leads to a different number of e-fold.}
    
\end{figure}

\begin{figure}
    \centering
    \includegraphics[width=8cm]{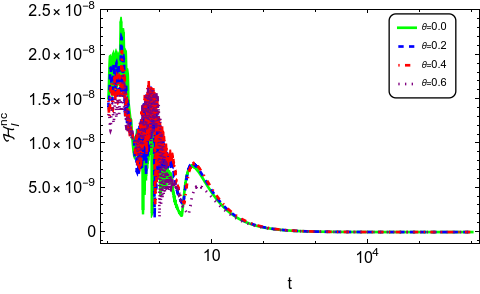}
    \caption{The figure shows the Hamiltonian constraint versus the cosmic time for different values of $\theta$. It is realized that the constraint is verified with a good approximation. }
    \label{c1Strb_Hamiltonian}
\end{figure}

The same subset of Eq.\eqref{shifted_variables} is applied to ensure simultaneous quantum bounce and maximum energy density. The same initial value of volume and the function $\beta$ are taken at the bounce, and the initial scalar field is taken as $\phi_B = -1.28$, which gives an inflation with enough number of e-folds in the commutative mLQC-I, and the initial value of the field momentum conjugate is achieved from the energy density at the bounce.    \\   
The numerical solutions obtained for the main phase space variables are illustrated in Fig.\eqref{c1Strb_main_quanities} for different values of the noncommutative parameter $\theta$. Same as in the chaotic potential case, the function $\beta$ and the scalar field are not affected by the noncommutativity parameter $\theta$ at the initial times around the bounce, and they overlap. The small differences appear at later times for these two functions. The other two variables are affected by the $\theta$; however, there is a different behavior for the volume and the field momentum conjugate. Here, by enhancing $\theta$, the universe expands less and the volume of the universe gets smaller. \\
%%%%%%%%%%%%%%%%%%%%%%%%%%%%%%%%%
%%%%%%%%%%%%%%%%%%%%%%%%%%%%%%%%%
It can also be seen in the behavior of the Hubble parameter and its time derivative, shown in Fig.\ref{c1Strb_HdH}. Unlike the chaotic case, the Hubble parameter is less for higher $\theta$, which shows a slower expansion rate. The behavior of $\dot{H}$ also indicates that the rate of $H$ is higher for smaller $\theta$. However, it drops faster so that the super-inflation stage is longer for a higher value of $\theta$. For instance, for $\theta = 0$, the super-inflation stage ends at $t = 0.38$, and it ends at $t = 0.61$ for $\theta = 0.6$. \\
%%%%%%%%%%%%%%%%%%%%%%%%%%%%%%%%%
%%%%%%%%%%%%%%%%%%%%%%%%%%%%%%%%% 
The change in the expansion rate of the universe also implies that the amount of its expansion is also changed. The number of e-folds versus the cosmic time is depicted in Fig.\eqref{c1Strb_efold_superinflation} for different values of $\theta$. It is realized that the initial expansion of the universe for a smaller value of $\theta$ is higher, and the universe experiences more e-fold expansion. 
%%%%%%%%%%%%%%%%%%%%%%%%%%%%%%%%% 
%%%%%%%%%%%%%%%%%%%%%%%%%%%%%%%%% 
Due to the behavior of the scalar field, the energy density is not expected to have a tangible change, especially around the bounce time. The plotted energy density of the scalar field, Fig.\ref{c1Strb_energy_density}, indicates that the curves overlap around the bounce point, and there would be a small variation for different values of $\theta$ at a later time. 
%%%%%%%%%%%%%%%%%%%%%%%%%%%%%%%%% 
%%%%%%%%%%%%%%%%%%%%%%%%%%%%%%%%% 
The same is expected for the equation of the state parameter $\omega$; displayed by  Fig.\ref{c1Strb_omega} for different values of $\theta$. It sounds like they are not affected by $\theta$ and overlap; however, there is a small variation at the end of the inflationary phase where the ones with higher $\theta$ stand for higher end time. 
%%%%%%%%%%%%%%%%%%%%%%%%%%%%%%%%% 
%%%%%%%%%%%%%%%%%%%%%%%%%%%%%%%%% 
This variation has a small effect on the total number of e-folds, as shown in Fig.\ref{c1Strb_efold_total}, so the differences in the total number of e-folds are about one. \\
At the end of the section, the Hamiltonian constraint is checked. The Fig.\ref{c1Strb_Hamiltonian} shows the Hamiltonian versus time for different values of $\theta$, which indicates that the constraint is verified during the time with good approximation. 
%%%%%%%%%%%%%%%%%%%%%%%%%%%%%%%%% 
%%%%%%%%%%%%%%%%%%%%%%%%%%%%%%%%%  

%%%%%%%%%%%%%%%%%%%%%%%%%%%%%%%%%%%%%%%%%%%%%%
%%%%%%%%%%%%%%%%%%%%%%%%%%%%%%%%%%%%%%%%%%%%%%
%%%%%%%%%%%%%%%%%%%%%%%%%%%%%%%%%%%%%%%%%%%%%%
%%%%%%%%%%%%%%%%%%%%%%%%%%%%%%%%%%%%%%%%%%%%%%
\subsection{Condition II: Chaotic potential}
\begin{figure}
    \centering
    \subfigure[$b = 1$]{\includegraphics[width=7cm]{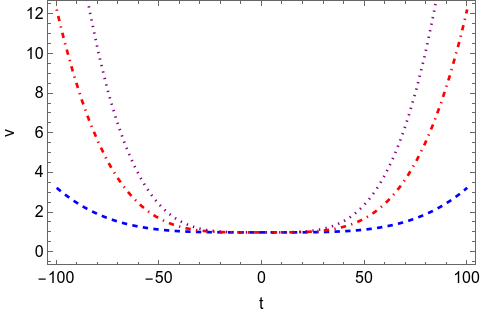}}
    \subfigure[$b = 10$]{\includegraphics[width=7cm]{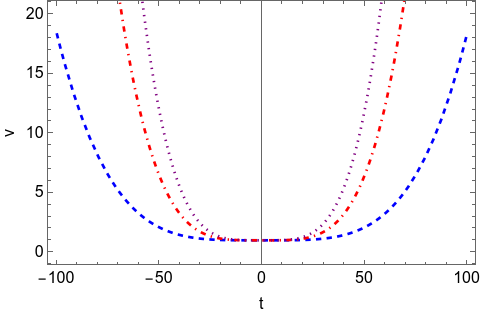}}
    \caption{The volume versus the cosmic time $t$ for different values of $\theta$ and a) $b = 1$ b) $b = 10$. The enhancement in the volume by increasing both $\theta$ and $b$ is clear from the plot. }
    \label{c2_v}
\end{figure}

\begin{figure}
    \centering
    \subfigure[$b = 1$]{\includegraphics[width=7cm]{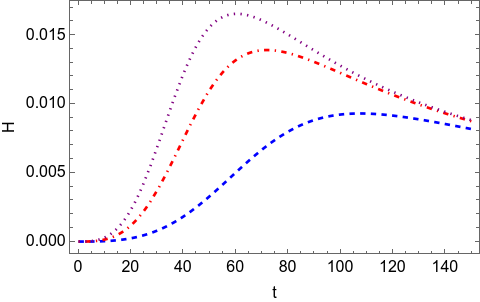}}
    \subfigure[$b = 10$]{\includegraphics[width=7cm]{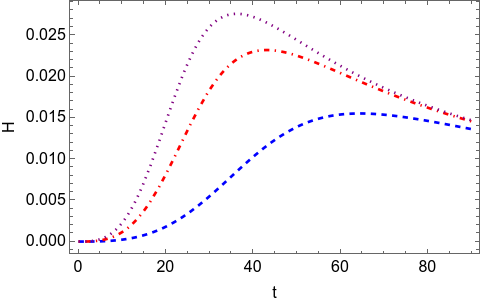}}
    \caption{The Hubble parameter versus the cosmic time $t$ for different values of $\theta$ and a) $b = 1$ b) $b = 10$. With an increase in $\theta$ and $b$, the Hubble parameter increases. Also, it has a higher peak that occurs in an earlier time. }
    \label{c2_H}
\end{figure}

\begin{figure}
    \centering
    \subfigure[$b = 1$]{\includegraphics[width=7cm]{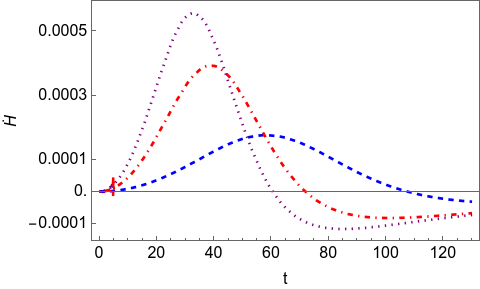}}
    \subfigure[$b = 10$]{\includegraphics[width=7cm]{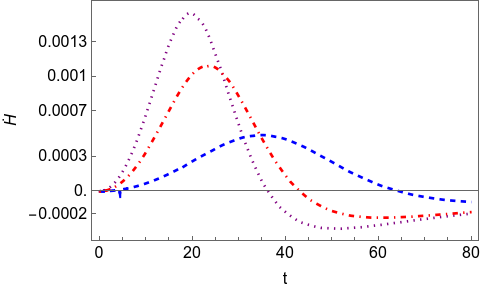}}
    \caption{The time derivative of the Hubble parameter $\dot{H}$ versus the cosmic time $t$ for different values of $\theta$ and a) $b = 1$ b) $b = 10$. For bigger $\theta$ and $b$, $\dot{H}$ first increase with higher rate and reaches sooner to a higher peak. Then it decreases with higher rate again and reach zero earlier. It means that there is shorter super-inflationary stage. }
    \label{c2_dH}
\end{figure}

The next condition we are going to consider is given as $\dot{\phi}(t_B) = 0$ and $b \theta \ddot{\phi} < 0$. This is another subset of Eq.\eqref{shifted_variables} that allows a bounce at the time when the energy density reaches the maximum value. Considering the Hamiltonian constraint, it indicates that the function $\beta$ should be $\beta(t_B) = \arcsin\big( 1 / \sqrt{2(\gamma^2 + 1)} \big) / \lambda$. On the other hand, applying the first statement of the condition to Eq.\eqref{phi1_equation} leads to the conclusion that at the bounce point the momentum conjugate of the scalar field is equal to $p_\phi(t_B) = - b \theta \; \beta(t_B) $. Since at the bounce $\dot\phi = 0$ and the energy density should be equal to the maximum value, the scalar field is confined as $\rho(t_B) = V(\phi_B) = \rho_c^I$. Scaling symmetry is used to set the initial value of the volume at the bounce as $v_B = 1$. Imposing these initial conditions on the differential equations Eqs.\eqref{v1_equation}, \eqref{b1_equation}, \eqref{phi1_equation}, and \eqref{pi1_equation}, the solutions are obtained numerically. Fig.\ref{c2_v} shows the volume versus the cosmic time for different values of the noncommutative parameters $\theta$ and $b$. It clearly shows that by enhancing the parameter $\theta$, there is a higher rate of expansion for the universe. Also, by increasing the parameter $b$, the universe experiences more expansion for the same time period. \\
%%%%%%%%%%%%%%%%%%%%%%%%%%%%%%%%%%%%%%%%%%%%
%%%%%%%%%%%%%%%%%%%%%%%%%%%%%%%%%%%%%%%%%%%% 
The expansion rate of the universe is determined by the Hubble parameter, which is plotted in Fig.\ref{c2_H}. As it was expected, the figure shows that the initial expansion rate of the universe increases with the enhancement of both $\theta$ and $b$; however, by passing time, one can see that the Hubble curves converge, and they describe almost the same expansion rate. Note that the maximum value of the Hubble parameter is greater for bigger values of $\theta$ and $b$. 
%%%%%%%%%%%%%%%%%%%%%%%%%%%%%%%%%%%%%%%%%%%%
%%%%%%%%%%%%%%%%%%%%%%%%%%%%%%%%%%%%%%%%%%%%  
In fact, at the initial time, the Hubble parameter increases with a bigger rate for higher values of $\theta$ and $\beta$, and reaches its maximum at an earlier time. This behavior can also be realized from Fig.\ref{c2_dH}, where $\dot{H}$ is plotted versus the cosmic time for different values of $\theta$ and $b$. It shows the same behavior for $\dot{H}$ as well. By enhancing $\theta$ and $b$, the time derivative of the Hubble parameter increases faster and reaches its maximum at an earlier time. Then, it decreases, and the decreasing rate is also faster, so that the stage of super-inflation finishes at a later time for smaller values of $\theta$ and $b$. 
%%%%%%%%%%%%%%%%%%%%%%%%%%%%%%%%%%%%%%%%%%%%
%%%%%%%%%%%%%%%%%%%%%%%%%%%%%%%%%%%%%%%%%%%%  

\begin{figure}
    \centering
    \subfigure[$b = 1$]{\includegraphics[width=7cm]{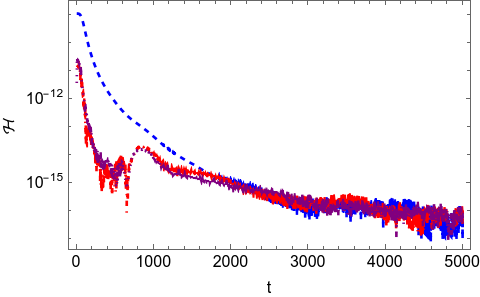}}
    \subfigure[$b = 10$]{\includegraphics[width=7cm]{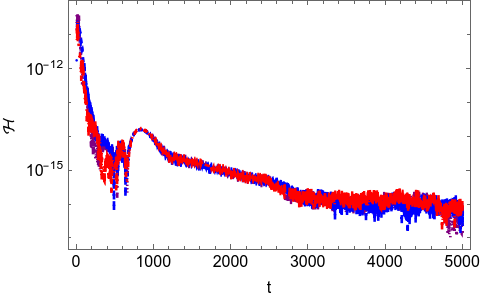}}
    \caption{The Hamiltonian constraint for two different choices of $b$ as a) $b = 1$ b) $b = 10$, and different values of the noncommutative parameter $\theta$ is plotted. In all cases, the Hamiltonian constraint is satisfied with a good approximation.  }
    \label{c2_hamiltonian}
\end{figure}

At the final step, we also consider the validity of the Hamiltonian constraint, which is plotted in Fig.\ref{c2_hamiltonian} for two choices of $b$ and different values of the noncommutative parameter $\theta$. It is realized that the numerical solution for the Hamiltonian is very close to zero, stating that the Hamiltonian constraint is satisfied with a good approximation. 
%%%%%%%%%%%%%%%%%%%%%%%%%%%%%%%%%%%%%%%%%%%%
%%%%%%%%%%%%%%%%%%%%%%%%%%%%%%%%%%%%%%%%%%%% 

%%%%%%%%%%%%%%%%%%%%%%%%%%%%%%%%%%%%%%%%%%%%%%
%%%%%%%%%%%%%%%%%%%%%%%%%%%%%%%%%%%%%%%%%%%%%%
%%%%%%%%%%%%%%%%%%%%%%%%%%%%%%%%%%%%%%%%%%%%%%
%%%%%%%%%%%%%%%%%%%%%%%%%%%%%%%%%%%%%%%%%%%%%%
\subsection{Condition II: Starobinsky potential}
\begin{figure}
    \centering
    \subfigure[$b = -1$]{\includegraphics[width=7cm]{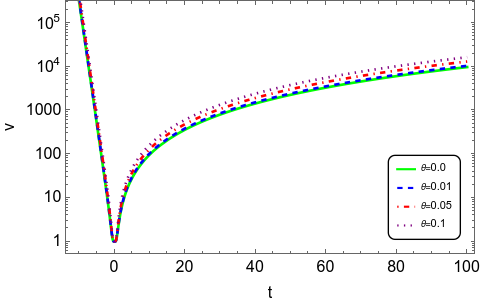}}
    \subfigure[$b = -10$]{\includegraphics[width=7cm]{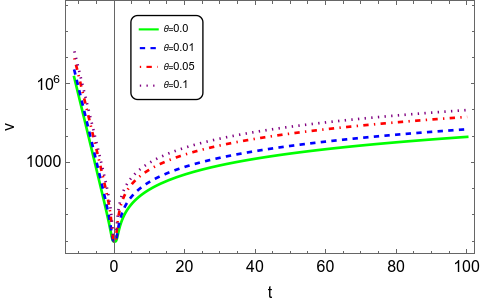}}
    \caption{The volume versus the cosmic time $t$ for different values of $\theta$ and a) $b = 1$ b) $b = 10$. Like the previous case, the expansion of the universe increases with an enhancement in $\theta$ and $b$. }
    \label{c2Strb_v}
\end{figure}

\begin{figure}
    \centering
    \subfigure[$b = -1$]{\includegraphics[width=7cm]{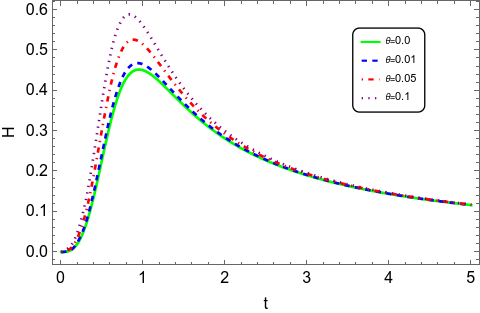}}
    \subfigure[$b = -10$]{\includegraphics[width=7cm]{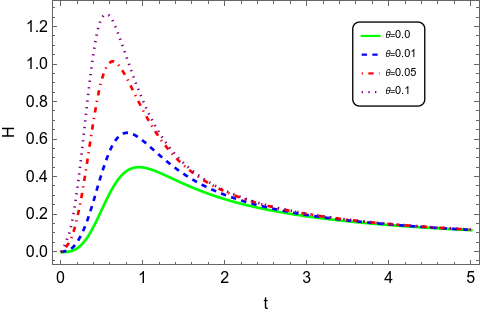}}
    \caption{The Hubble parameter versus the cosmic time $t$ for different values of $\theta$ and a) $b = 1$ b) $b = 10$.   }
    \label{c2Strb_H}
\end{figure}

\begin{figure}
    \centering
    \subfigure[$b = -1$]{\includegraphics[width=7cm]{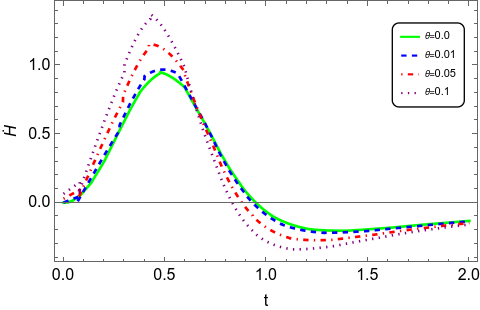}}
    \subfigure[$b = -10$]{\includegraphics[width=7cm]{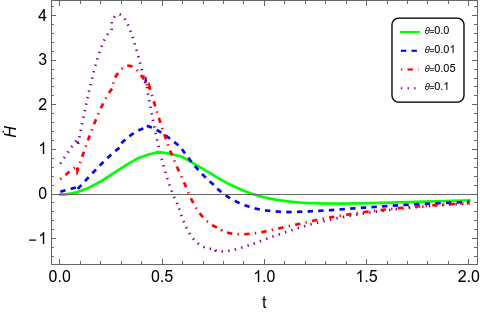}}
    \caption{The time derivative of the Hubble parameter $\dot{H}$ versus the cosmic time $t$ for different values of $\theta$ and a) $b = 1$ b) $b = 10$. Higher $\theta$ and $b$ leads to bigger initial increasing rate and secondary decreasing rate. The super-inflationary stage ends earlier.  }
    \label{c2Strb_dH}
\end{figure}

\begin{figure}
    \centering
    \subfigure[$b = -1$]{\includegraphics[width=7cm]{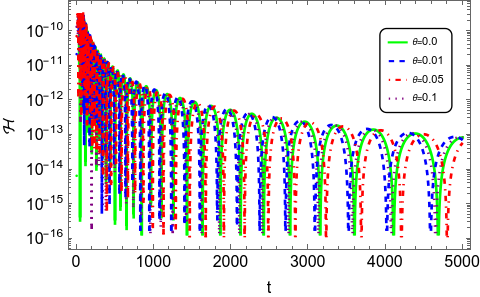}}
    \subfigure[$b = -10$]{\includegraphics[width=7cm]{fig1NC22_Hamiltonianb10_Strb.png}}
    \caption{The Hamiltonian constraint for two different choices of $b$ as a) $b = 1$ b) $b = 10$, and different values of the noncommutative parameter $\theta$ is illustrated, indicating that the constraint is satisfied with a good approximation. }
    \label{c2Strb_hamiltonian}
\end{figure}

Considering the Starobinsky potential, we are going to study the evolution of the universe for the second subset of \eqref{shifted_variables}. To verify this condition, as mentioned in the previous section, there should be $b \theta \ddot\phi(t_B) < 0$. Since the derivative of the Starobinsky potential is negative, the constants $b$ and $\theta$ should be chosen so that $b \theta < 0$. Here, we keep $\theta$ to be positive, and we choose $b$ the same value with the negative sign. \\
Pursuing a numerical approach and solving the equation, the solution for the volume shows dependency on $\theta$ so that the universe expands more for higher $\theta$, and it gets even higher by increasing the magnitude of $b$. It can be seen in Fig.\ref{c2Strb_v}.  \\  
%%%%%%%%%%%%%%%%%%%%%%%%%%%%%%%%%%%%%%%%%%%%
%%%%%%%%%%%%%%%%%%%%%%%%%%%%%%%%%%%%%%%%%%%% 
The increasing rate of the universe is understood through the behavior of the Hubble parameter, plotted in Fig.\ref{c2Strb_H}. It clearly displays that by increasing the value of $\theta$, the Hubble parameter increases and reaches a peak of higher magnitude. This indicates that the universe expansion rate is higher for larger $\theta$, and it even increases more with an enhancement of the magnitude of $b$. However, the different Hubble curves finally converge. Comparing to the chaotic potential, the magnitude of the Hubble parameter is higher, and it reaches its maximum at an earlier time. 
%%%%%%%%%%%%%%%%%%%%%%%%%%%%%%%%%%%%%%%%%%%%
%%%%%%%%%%%%%%%%%%%%%%%%%%%%%%%%%%%%%%%%%%%% 
Considering the $\dot{H}$ in Fig.\ref{c2Strb_dH} also exhibits that the enhancement of the Hubble parameter is higher for bigger values of $\theta$ and $b$, and their decreasing rate is also faster. $\dot{H}$ is positive initially and reaches zero at an earlier time by increasing $\theta$ and $b$, implying a shorter super-inflationary stage. In comparison to the chaotic potential, the initial increasing rate and the second decreasing rate of $\dot{H}$ are bigger, and the time period of the super-inflationary stage is shorter. \\   
%%%%%%%%%%%%%%%%%%%%%%%%%%%%%%%%%%%%%%%%%%%%
%%%%%%%%%%%%%%%%%%%%%%%%%%%%%%%%%%%%%%%%%%%%  
Fig.\ref{c2Strb_hamiltonian} displays the Hamiltonian constraint versus the cosmic time for different values of $\theta$ and $b$, and with a good approximation, it can be said that the constraint is verified during the time. 
%%%%%%%%%%%%%%%%%%%%%%%%%%%%%%%%%%%%%%%%%%%%
%%%%%%%%%%%%%%%%%%%%%%%%%%%%%%%%%%%%%%%%%%%% 

%%%%%%%%%%%%%%%%%%%%%%%%%%%%%%%%%%%%%%%%%%%%%%
%%%%%%%%%%%%%%%%%%%%%%%%%%%%%%%%%%%%%%%%%%%%%%
%%%%%%%%%%%%%%%%%%%%%%%%%%%%%%%%%%%%%%%%%%%%%%
%%%%%%%%%%%%%%%%%%%%%%%%%%%%%%%%%%%%%%%%%%%%%%
%%%%%%%%%%%%%%%%%%%%%%%%%%%%%%%%%%%%%%%%%%%%%%
%%%%%%%%%%%%%%%%%%%%%%%%%%%%%%%%%%%%%%%%%%%%%%
\section{Conclusion}
Considering a simple noncommutative extension of the modified LQC, the pre-inflationary dynamics of the universe  was considered. Among the two approaches for formulating the deformed algebra, the shifted variables prescription was chosen where the resulted Hamiltonian has the same functional form as the commutative one. The deformed algebra was considered in the momentum sector, and after constructing the effective Hamiltonian, the modified equations were obtained for the phase space variables $\{ \beta, \phi, v, p_\phi \}$. The model was studied for chaotic and Starobinsky potentials. \\
Interestingly, the equation for the function $\beta$ has the same form as the one in commutative mLQC-I, however, the equations for the rest of the variables are modified. The energy density is also in the same functional form in terms of $\beta$, implying that there is a maximum value for the energy density. The complexity of the equations makes it difficult to get to the analytical solution, so a numerical approach was followed. Moreover, due to the modification in the $v$ equation, it can not be guaranteed in general that the quantum bounce, where $\dot{v} = 0$, and the maximum of the energy density appear simultaneously. To align these events, two sets of conditions were imposed and analyzed separately. The first set of conditions is $b = 0$ and $v^{nc} > 0$, so that the commutative mLQC-I volume equations is restored. This ensures that when the energy density reaches the maximum value $\rho_c^I$ there is a quantum bounce. The noncommutativity in this case changes the behavior of the volume, so that, by enhancement of $\theta$, the volume gets bigger when the potential is taken to be chaotic potential; however, the volume decreases for Starobinsky potential. It also affects the expansion rate of the universe. As shown in Figs.\ref{c1_HdH} and \ref{c1Strb_HdH}, it was found that by considering the chaotic potential for the scalar field, the Hubble parameter and its time derivative increase, which leads to a higher e-fold of expansion for the universe and a shorter superinflationary stage. Conversely, the Starobinsky potential exhibited opposite behavior, in which, with an increase in the $\theta$ parameter the Hubble parameter and its time derivative decrease resulted in a less e-fold of expansion for the universe and a longer super-inflationary stage, shown in Fig.\ref{c1Strb_HdH}. The behavior of the scalar field $\phi$ and $\beta$ are slightly affected by the variation of $\theta$ at later times, while they overlap at times around the bounce. This leads to the similar behavior for the energy density. The transition to the inflationary phase happens almost at the same time. The total number of e-folds, from the quantum bounce to the end of the inflationary phase, is a little higher/lower for a bigger value of $\theta$ in the case of chaotic/Starobinsky potential. \\    
The second set of conditions is $\dot\phi(t_c) = 0$ and $b \theta \ddot{\phi} < 0$. This condition also provides a sufficient condition for a big bounce, although more stationary points are not excluded by this condition, and it might lead to several bounce points. The initial values of the scalar field and its momentum conjugate are also determined by this condition. Numerical solution shows, in general, the same behavior for both potentials. Here, it is also determined that by enhancing the parameters $b$ and $\theta$, the volume of the universe gets bigger. The universe expansion rate is effectively affected by $b$ and $\theta$. With increasing these parameters, the Hubble parameter's peak is located higher, reached faster, and then it is declines more rapidly that results in a shorter superinflationary period compared to commutative mLQC-I. \\
The Hamiltonian constraint, as one of the fundamentals of the theory that ensures the consistency of the equations, was considered for both cases, and the result showed that the constraint is verified with a good approximation.  \\

%\nocite{*}

\bibliography{RefBib}% Produces the bibliography via BibTeX.

\begin{thebibliography}{58}%
\makeatletter
\providecommand \@ifxundefined [1]{%
 \@ifx{#1\undefined}
}%
\providecommand \@ifnum [1]{%
 \ifnum #1\expandafter \@firstoftwo
 \else \expandafter \@secondoftwo
 \fi
}%
\providecommand \@ifx [1]{%
 \ifx #1\expandafter \@firstoftwo
 \else \expandafter \@secondoftwo
 \fi
}%
\providecommand \natexlab [1]{#1}%
\providecommand \enquote  [1]{``#1''}%
\providecommand \bibnamefont  [1]{#1}%
\providecommand \bibfnamefont [1]{#1}%
\providecommand \citenamefont [1]{#1}%
\providecommand \href@noop [0]{\@secondoftwo}%
\providecommand \href [0]{\begingroup \@sanitize@url \@href}%
\providecommand \@href[1]{\@@startlink{#1}\@@href}%
\providecommand \@@href[1]{\endgroup#1\@@endlink}%
\providecommand \@sanitize@url [0]{\catcode `\\12\catcode `\$12\catcode
  `\&12\catcode `\#12\catcode `\^12\catcode `\_12\catcode `\%12\relax}%
\providecommand \@@startlink[1]{}%
\providecommand \@@endlink[0]{}%
\providecommand \url  [0]{\begingroup\@sanitize@url \@url }%
\providecommand \@url [1]{\endgroup\@href {#1}{\urlprefix }}%
\providecommand \urlprefix  [0]{URL }%
\providecommand \Eprint [0]{\href }%
\providecommand \doibase [0]{https://doi.org/}%
\providecommand \selectlanguage [0]{\@gobble}%
\providecommand \bibinfo  [0]{\@secondoftwo}%
\providecommand \bibfield  [0]{\@secondoftwo}%
\providecommand \translation [1]{[#1]}%
\providecommand \BibitemOpen [0]{}%
\providecommand \bibitemStop [0]{}%
\providecommand \bibitemNoStop [0]{.\EOS\space}%
\providecommand \EOS [0]{\spacefactor3000\relax}%
\providecommand \BibitemShut  [1]{\csname bibitem#1\endcsname}%
\let\auto@bib@innerbib\@empty
%</preamble>
\bibitem [{\citenamefont {Rovelli}(2004)}]{rovelli2004quantum}%
  \BibitemOpen
  \bibfield  {author} {\bibinfo {author} {\bibfnamefont {C.}~\bibnamefont
  {Rovelli}},\ }\href@noop {} {\emph {\bibinfo {title} {Quantum gravity}}}\
  (\bibinfo  {publisher} {Cambridge university press},\ \bibinfo {year}
  {2004})\BibitemShut {NoStop}%
\bibitem [{\citenamefont {Thiemann}(2008)}]{thiemann2008modern}%
  \BibitemOpen
  \bibfield  {author} {\bibinfo {author} {\bibfnamefont {T.}~\bibnamefont
  {Thiemann}},\ }\href@noop {} {\emph {\bibinfo {title} {Modern canonical
  quantum general relativity}}}\ (\bibinfo  {publisher} {Cambridge University
  Press},\ \bibinfo {year} {2008})\BibitemShut {NoStop}%
\bibitem [{\citenamefont {Singh}(2009)}]{Singh:2009mz}%
  \BibitemOpen
  \bibfield  {author} {\bibinfo {author} {\bibfnamefont {P.}~\bibnamefont
  {Singh}},\ }\bibfield  {title} {\bibinfo {title} {{Are loop quantum cosmos
  never singular?}},\ }\href {https://doi.org/10.1088/0264-9381/26/12/125005}
  {\bibfield  {journal} {\bibinfo  {journal} {Class. Quant. Grav.}\ }\textbf
  {\bibinfo {volume} {26}},\ \bibinfo {pages} {125005} (\bibinfo {year}
  {2009})},\ \Eprint {https://arxiv.org/abs/0901.2750} {arXiv:0901.2750
  [gr-qc]} \BibitemShut {NoStop}%
\bibitem [{\citenamefont {Singh}(2014)}]{Singh:2014fsy}%
  \BibitemOpen
  \bibfield  {author} {\bibinfo {author} {\bibfnamefont {P.}~\bibnamefont
  {Singh}},\ }\bibfield  {title} {\bibinfo {title} {{Loop quantum cosmology and
  the fate of cosmological singularities}},\ }\href@noop {} {\bibfield
  {journal} {\bibinfo  {journal} {Bull. Astron. Soc. India}\ }\textbf {\bibinfo
  {volume} {42}},\ \bibinfo {pages} {121} (\bibinfo {year} {2014})},\ \Eprint
  {https://arxiv.org/abs/1509.09182} {arXiv:1509.09182 [gr-qc]} \BibitemShut
  {NoStop}%
\bibitem [{\citenamefont {Snyder}(1947)}]{Snyder:1946qz}%
  \BibitemOpen
  \bibfield  {author} {\bibinfo {author} {\bibfnamefont {H.~S.}\ \bibnamefont
  {Snyder}},\ }\bibfield  {title} {\bibinfo {title} {{Quantized space-time}},\
  }\href {https://doi.org/10.1103/PhysRev.71.38} {\bibfield  {journal}
  {\bibinfo  {journal} {Phys. Rev.}\ }\textbf {\bibinfo {volume} {71}},\
  \bibinfo {pages} {38} (\bibinfo {year} {1947})}\BibitemShut {NoStop}%
\bibitem [{\citenamefont {Connes}(1994)}]{Connes:1994yd}%
  \BibitemOpen
  \bibfield  {author} {\bibinfo {author} {\bibfnamefont {A.}~\bibnamefont
  {Connes}},\ }\href@noop {} {\emph {\bibinfo {title} {{Noncommutative
  geometry}}}}\ (\bibinfo  {publisher} {Academic Press},\ \bibinfo {year}
  {1994})\BibitemShut {NoStop}%
\bibitem [{\citenamefont {Lee}\ \emph {et~al.}(1998)\citenamefont {Lee},
  \citenamefont {Leung},\ and\ \citenamefont {Ng}}]{Lee:1997uh}%
  \BibitemOpen
  \bibfield  {author} {\bibinfo {author} {\bibfnamefont {D.~S.}\ \bibnamefont
  {Lee}}, \bibinfo {author} {\bibfnamefont {C.~N.}\ \bibnamefont {Leung}},\
  and\ \bibinfo {author} {\bibfnamefont {Y.~J.}\ \bibnamefont {Ng}},\
  }\bibfield  {title} {\bibinfo {title} {{Chiral symmetry breaking in a uniform
  external magnetic field. 2. Symmetry restoration at high temperatures and
  chemical potentials}},\ }\href {https://doi.org/10.1103/PhysRevD.57.5224}
  {\bibfield  {journal} {\bibinfo  {journal} {Phys. Rev. D}\ }\textbf {\bibinfo
  {volume} {57}},\ \bibinfo {pages} {5224} (\bibinfo {year} {1998})},\ \Eprint
  {https://arxiv.org/abs/hep-th/9711126} {arXiv:hep-th/9711126} \BibitemShut
  {NoStop}%
\bibitem [{\citenamefont {Seiberg}\ and\ \citenamefont
  {Witten}(1999)}]{Seiberg:1999vs}%
  \BibitemOpen
  \bibfield  {author} {\bibinfo {author} {\bibfnamefont {N.}~\bibnamefont
  {Seiberg}}\ and\ \bibinfo {author} {\bibfnamefont {E.}~\bibnamefont
  {Witten}},\ }\bibfield  {title} {\bibinfo {title} {{String theory and
  noncommutative geometry}},\ }\href
  {https://doi.org/10.1088/1126-6708/1999/09/032} {\bibfield  {journal}
  {\bibinfo  {journal} {JHEP}\ }\textbf {\bibinfo {volume} {09}},\ \bibinfo
  {pages} {032}},\ \Eprint {https://arxiv.org/abs/hep-th/9908142}
  {arXiv:hep-th/9908142} \BibitemShut {NoStop}%
\bibitem [{\citenamefont {Connes}\ \emph {et~al.}(1998)\citenamefont {Connes},
  \citenamefont {Douglas},\ and\ \citenamefont {Schwarz}}]{Connes:1997cr}%
  \BibitemOpen
  \bibfield  {author} {\bibinfo {author} {\bibfnamefont {A.}~\bibnamefont
  {Connes}}, \bibinfo {author} {\bibfnamefont {M.~R.}\ \bibnamefont
  {Douglas}},\ and\ \bibinfo {author} {\bibfnamefont {A.~S.}\ \bibnamefont
  {Schwarz}},\ }\bibfield  {title} {\bibinfo {title} {{Noncommutative geometry
  and matrix theory: Compactification on tori}},\ }\href
  {https://doi.org/10.1088/1126-6708/1998/02/003} {\bibfield  {journal}
  {\bibinfo  {journal} {JHEP}\ }\textbf {\bibinfo {volume} {02}},\ \bibinfo
  {pages} {003}},\ \Eprint {https://arxiv.org/abs/hep-th/9711162}
  {arXiv:hep-th/9711162} \BibitemShut {NoStop}%
\bibitem [{\citenamefont {Chu}\ and\ \citenamefont {Ho}(1999)}]{Chu:1998qz}%
  \BibitemOpen
  \bibfield  {author} {\bibinfo {author} {\bibfnamefont {C.-S.}\ \bibnamefont
  {Chu}}\ and\ \bibinfo {author} {\bibfnamefont {P.-M.}\ \bibnamefont {Ho}},\
  }\bibfield  {title} {\bibinfo {title} {{Noncommutative open string and
  D-brane}},\ }\href {https://doi.org/10.1016/S0550-3213(99)00199-6} {\bibfield
   {journal} {\bibinfo  {journal} {Nucl. Phys. B}\ }\textbf {\bibinfo {volume}
  {550}},\ \bibinfo {pages} {151} (\bibinfo {year} {1999})},\ \Eprint
  {https://arxiv.org/abs/hep-th/9812219} {arXiv:hep-th/9812219} \BibitemShut
  {NoStop}%
\bibitem [{\citenamefont {Schomerus}(1999)}]{Schomerus:1999ug}%
  \BibitemOpen
  \bibfield  {author} {\bibinfo {author} {\bibfnamefont {V.}~\bibnamefont
  {Schomerus}},\ }\bibfield  {title} {\bibinfo {title} {{D-branes and
  deformation quantization}},\ }\href
  {https://doi.org/10.1088/1126-6708/1999/06/030} {\bibfield  {journal}
  {\bibinfo  {journal} {JHEP}\ }\textbf {\bibinfo {volume} {06}},\ \bibinfo
  {pages} {030}},\ \Eprint {https://arxiv.org/abs/hep-th/9903205}
  {arXiv:hep-th/9903205} \BibitemShut {NoStop}%
\bibitem [{\citenamefont {Garcia-Compean}\ \emph {et~al.}(2002)\citenamefont
  {Garcia-Compean}, \citenamefont {Obregon},\ and\ \citenamefont
  {Ramirez}}]{Garcia-Compean:2001jxk}%
  \BibitemOpen
  \bibfield  {author} {\bibinfo {author} {\bibfnamefont {H.}~\bibnamefont
  {Garcia-Compean}}, \bibinfo {author} {\bibfnamefont {O.}~\bibnamefont
  {Obregon}},\ and\ \bibinfo {author} {\bibfnamefont {C.}~\bibnamefont
  {Ramirez}},\ }\bibfield  {title} {\bibinfo {title} {{Noncommutative quantum
  cosmology}},\ }\href {https://doi.org/10.1103/PhysRevLett.88.161301}
  {\bibfield  {journal} {\bibinfo  {journal} {Phys. Rev. Lett.}\ }\textbf
  {\bibinfo {volume} {88}},\ \bibinfo {pages} {161301} (\bibinfo {year}
  {2002})},\ \Eprint {https://arxiv.org/abs/hep-th/0107250}
  {arXiv:hep-th/0107250} \BibitemShut {NoStop}%
\bibitem [{\citenamefont {Douglas}\ and\ \citenamefont
  {Nekrasov}(2001)}]{Douglas:2001ba}%
  \BibitemOpen
  \bibfield  {author} {\bibinfo {author} {\bibfnamefont {M.~R.}\ \bibnamefont
  {Douglas}}\ and\ \bibinfo {author} {\bibfnamefont {N.~A.}\ \bibnamefont
  {Nekrasov}},\ }\bibfield  {title} {\bibinfo {title} {{Noncommutative field
  theory}},\ }\href {https://doi.org/10.1103/RevModPhys.73.977} {\bibfield
  {journal} {\bibinfo  {journal} {Rev. Mod. Phys.}\ }\textbf {\bibinfo {volume}
  {73}},\ \bibinfo {pages} {977} (\bibinfo {year} {2001})},\ \Eprint
  {https://arxiv.org/abs/hep-th/0106048} {arXiv:hep-th/0106048} \BibitemShut
  {NoStop}%
\bibitem [{\citenamefont {Szabo}(2003)}]{Szabo:2001kg}%
  \BibitemOpen
  \bibfield  {author} {\bibinfo {author} {\bibfnamefont {R.~J.}\ \bibnamefont
  {Szabo}},\ }\bibfield  {title} {\bibinfo {title} {{Quantum field theory on
  noncommutative spaces}},\ }\href
  {https://doi.org/10.1016/S0370-1573(03)00059-0} {\bibfield  {journal}
  {\bibinfo  {journal} {Phys. Rept.}\ }\textbf {\bibinfo {volume} {378}},\
  \bibinfo {pages} {207} (\bibinfo {year} {2003})},\ \Eprint
  {https://arxiv.org/abs/hep-th/0109162} {arXiv:hep-th/0109162} \BibitemShut
  {NoStop}%
\bibitem [{\citenamefont {Moffat}(2000)}]{Moffat:2000gr}%
  \BibitemOpen
  \bibfield  {author} {\bibinfo {author} {\bibfnamefont {J.~W.}\ \bibnamefont
  {Moffat}},\ }\bibfield  {title} {\bibinfo {title} {{Noncommutative quantum
  gravity}},\ }\href {https://doi.org/10.1016/S0370-2693(00)01064-9} {\bibfield
   {journal} {\bibinfo  {journal} {Phys. Lett. B}\ }\textbf {\bibinfo {volume}
  {491}},\ \bibinfo {pages} {345} (\bibinfo {year} {2000})},\ \Eprint
  {https://arxiv.org/abs/hep-th/0007181} {arXiv:hep-th/0007181} \BibitemShut
  {NoStop}%
\bibitem [{\citenamefont {Chamseddine}(2001)}]{Chamseddine:2000zu}%
  \BibitemOpen
  \bibfield  {author} {\bibinfo {author} {\bibfnamefont {A.~H.}\ \bibnamefont
  {Chamseddine}},\ }\bibfield  {title} {\bibinfo {title} {{Complexified gravity
  in noncommutative spaces}},\ }\href {https://doi.org/10.1007/s002200100393}
  {\bibfield  {journal} {\bibinfo  {journal} {Commun. Math. Phys.}\ }\textbf
  {\bibinfo {volume} {218}},\ \bibinfo {pages} {283} (\bibinfo {year}
  {2001})},\ \Eprint {https://arxiv.org/abs/hep-th/0005222}
  {arXiv:hep-th/0005222} \BibitemShut {NoStop}%
\bibitem [{\citenamefont {Garcia-Compean}\ \emph {et~al.}(2003)\citenamefont
  {Garcia-Compean}, \citenamefont {Obregon}, \citenamefont {Ramirez},\ and\
  \citenamefont {Sabido}}]{Garcia-Compean:2003nix}%
  \BibitemOpen
  \bibfield  {author} {\bibinfo {author} {\bibfnamefont {H.}~\bibnamefont
  {Garcia-Compean}}, \bibinfo {author} {\bibfnamefont {O.}~\bibnamefont
  {Obregon}}, \bibinfo {author} {\bibfnamefont {C.}~\bibnamefont {Ramirez}},\
  and\ \bibinfo {author} {\bibfnamefont {M.}~\bibnamefont {Sabido}},\
  }\bibfield  {title} {\bibinfo {title} {{Noncommutative selfdual gravity}},\
  }\href {https://doi.org/10.1103/PhysRevD.68.044015} {\bibfield  {journal}
  {\bibinfo  {journal} {Phys. Rev. D}\ }\textbf {\bibinfo {volume} {68}},\
  \bibinfo {pages} {044015} (\bibinfo {year} {2003})},\ \Eprint
  {https://arxiv.org/abs/hep-th/0302180} {arXiv:hep-th/0302180} \BibitemShut
  {NoStop}%
\bibitem [{\citenamefont {Kober}(2015)}]{Kober:2014wsa}%
  \BibitemOpen
  \bibfield  {author} {\bibinfo {author} {\bibfnamefont {M.}~\bibnamefont
  {Kober}},\ }\bibfield  {title} {\bibinfo {title} {{Canonical quantum gravity
  on noncommutative space\textendash{}time}},\ }\href
  {https://doi.org/10.1142/S0217751X15500852} {\bibfield  {journal} {\bibinfo
  {journal} {Int. J. Mod. Phys. A}\ }\textbf {\bibinfo {volume} {30}},\
  \bibinfo {pages} {1550085} (\bibinfo {year} {2015})},\ \Eprint
  {https://arxiv.org/abs/1409.1751} {arXiv:1409.1751 [gr-qc]} \BibitemShut
  {NoStop}%
\bibitem [{\citenamefont {Espinoza-Garc\'\i{}a}\ \emph
  {et~al.}(2019)\citenamefont {Espinoza-Garc\'\i{}a}, \citenamefont
  {Torres-Lomas}, \citenamefont {P\'erez-Pay\'an},\ and\ \citenamefont
  {D\'\i{}az-Barr\'on}}]{Espinoza-Garcia:2017qjl}%
  \BibitemOpen
  \bibfield  {author} {\bibinfo {author} {\bibfnamefont {A.}~\bibnamefont
  {Espinoza-Garc\'\i{}a}}, \bibinfo {author} {\bibfnamefont {E.}~\bibnamefont
  {Torres-Lomas}}, \bibinfo {author} {\bibfnamefont {S.}~\bibnamefont
  {P\'erez-Pay\'an}},\ and\ \bibinfo {author} {\bibfnamefont {L.~R.}\
  \bibnamefont {D\'\i{}az-Barr\'on}},\ }\bibfield  {title} {\bibinfo {title}
  {{Noncommutativity in Effective Loop Quantum Cosmology}},\ }\href
  {https://doi.org/10.1155/2019/9080218} {\bibfield  {journal} {\bibinfo
  {journal} {Adv. High Energy Phys.}\ }\textbf {\bibinfo {volume} {2019}},\
  \bibinfo {pages} {9080218} (\bibinfo {year} {2019})},\ \Eprint
  {https://arxiv.org/abs/1709.03242} {arXiv:1709.03242 [gr-qc]} \BibitemShut
  {NoStop}%
\bibitem [{\citenamefont {D\'\i{}az-Barr\'on}\ \emph
  {et~al.}(2020)\citenamefont {D\'\i{}az-Barr\'on}, \citenamefont
  {Espinoza-Garc\'\i{}a}, \citenamefont {P\'erez-Pay\'an},\ and\ \citenamefont
  {Socorro}}]{Diaz-Barron:2019awc}%
  \BibitemOpen
  \bibfield  {author} {\bibinfo {author} {\bibfnamefont {L.~R.}\ \bibnamefont
  {D\'\i{}az-Barr\'on}}, \bibinfo {author} {\bibfnamefont {A.}~\bibnamefont
  {Espinoza-Garc\'\i{}a}}, \bibinfo {author} {\bibfnamefont {S.}~\bibnamefont
  {P\'erez-Pay\'an}},\ and\ \bibinfo {author} {\bibfnamefont {J.}~\bibnamefont
  {Socorro}},\ }\bibfield  {title} {\bibinfo {title} {{Noncommutative Friedmann
  Equations in Effective LQC}},\ }\href
  {https://doi.org/10.1142/S021827182050039X} {\bibfield  {journal} {\bibinfo
  {journal} {Int. J. Mod. Phys. D}\ }\textbf {\bibinfo {volume} {29}},\
  \bibinfo {pages} {2050039} (\bibinfo {year} {2020})},\ \Eprint
  {https://arxiv.org/abs/1904.01212} {arXiv:1904.01212 [gr-qc]} \BibitemShut
  {NoStop}%
\bibitem [{\citenamefont {D\'\i{}az-Barr\'on}\ \emph
  {et~al.}(2021)\citenamefont {D\'\i{}az-Barr\'on}, \citenamefont
  {Espinoza-Garc\'\i{}a}, \citenamefont {P\'erez-Pay\'an},\ and\ \citenamefont
  {Socorro}}]{Diaz-Barron:2021yha}%
  \BibitemOpen
  \bibfield  {author} {\bibinfo {author} {\bibfnamefont {L.~R.}\ \bibnamefont
  {D\'\i{}az-Barr\'on}}, \bibinfo {author} {\bibfnamefont {A.}~\bibnamefont
  {Espinoza-Garc\'\i{}a}}, \bibinfo {author} {\bibfnamefont {S.}~\bibnamefont
  {P\'erez-Pay\'an}},\ and\ \bibinfo {author} {\bibfnamefont {J.}~\bibnamefont
  {Socorro}},\ }\bibfield  {title} {\bibinfo {title} {{Noncommutative effective
  loop quantum cosmology: Inclusion of a potential term}},\ }\href
  {https://doi.org/10.1103/PhysRevD.104.023508} {\bibfield  {journal} {\bibinfo
   {journal} {Phys. Rev. D}\ }\textbf {\bibinfo {volume} {104}},\ \bibinfo
  {pages} {023508} (\bibinfo {year} {2021})},\ \Eprint
  {https://arxiv.org/abs/2104.02859} {arXiv:2104.02859 [gr-qc]} \BibitemShut
  {NoStop}%
\bibitem [{\citenamefont {D\'\i{}az-Barr\'on}\ \emph
  {et~al.}(2023{\natexlab{a}})\citenamefont {D\'\i{}az-Barr\'on}, \citenamefont
  {Espinoza-Garc\'\i{}a}, \citenamefont {P\'erez-Pay\'an},\ and\ \citenamefont
  {Socorro}}]{Diaz-Barron:2023ctp}%
  \BibitemOpen
  \bibfield  {author} {\bibinfo {author} {\bibfnamefont {L.~R.}\ \bibnamefont
  {D\'\i{}az-Barr\'on}}, \bibinfo {author} {\bibfnamefont {A.}~\bibnamefont
  {Espinoza-Garc\'\i{}a}}, \bibinfo {author} {\bibfnamefont {S.}~\bibnamefont
  {P\'erez-Pay\'an}},\ and\ \bibinfo {author} {\bibfnamefont {J.}~\bibnamefont
  {Socorro}},\ }\bibfield  {title} {\bibinfo {title} {{Is a loopy and
  noncommutative early Universe viable?}},\ }\href
  {https://doi.org/10.1016/j.physletb.2023.138299} {\bibfield  {journal}
  {\bibinfo  {journal} {Phys. Lett. B}\ }\textbf {\bibinfo {volume} {847}},\
  \bibinfo {pages} {138299} (\bibinfo {year} {2023}{\natexlab{a}})}\BibitemShut
  {NoStop}%
\bibitem [{\citenamefont {D\'\i{}az-Barr\'on}\ \emph
  {et~al.}(2023{\natexlab{b}})\citenamefont {D\'\i{}az-Barr\'on}, \citenamefont
  {Espinoza-Garc\'\i{}a}, \citenamefont {P\'erez-Pay\'an},\ and\ \citenamefont
  {Socorro}}]{Diaz-Barron:2023qcf}%
  \BibitemOpen
  \bibfield  {author} {\bibinfo {author} {\bibfnamefont {L.~R.}\ \bibnamefont
  {D\'\i{}az-Barr\'on}}, \bibinfo {author} {\bibfnamefont {A.}~\bibnamefont
  {Espinoza-Garc\'\i{}a}}, \bibinfo {author} {\bibfnamefont {S.}~\bibnamefont
  {P\'erez-Pay\'an}},\ and\ \bibinfo {author} {\bibfnamefont {J.}~\bibnamefont
  {Socorro}},\ }\bibfield  {title} {\bibinfo {title} {{Noncommutative effective
  LQC: A (pre-)inflationary dynamics investigation}},\ }\href@noop {} {\
  (\bibinfo {year} {2023}{\natexlab{b}})},\ \Eprint
  {https://arxiv.org/abs/2308.01363} {arXiv:2308.01363 [gr-qc]} \BibitemShut
  {NoStop}%
\bibitem [{\citenamefont {Thiemann}(1998{\natexlab{a}})}]{Thiemann:1996aw}%
  \BibitemOpen
  \bibfield  {author} {\bibinfo {author} {\bibfnamefont {T.}~\bibnamefont
  {Thiemann}},\ }\bibfield  {title} {\bibinfo {title} {{Quantum spin dynamics
  (QSD)}},\ }\href {https://doi.org/10.1088/0264-9381/15/4/011} {\bibfield
  {journal} {\bibinfo  {journal} {Class. Quant. Grav.}\ }\textbf {\bibinfo
  {volume} {15}},\ \bibinfo {pages} {839} (\bibinfo {year}
  {1998}{\natexlab{a}})},\ \Eprint {https://arxiv.org/abs/gr-qc/9606089}
  {arXiv:gr-qc/9606089} \BibitemShut {NoStop}%
\bibitem [{\citenamefont {Thiemann}(1998{\natexlab{b}})}]{Thiemann:1996av}%
  \BibitemOpen
  \bibfield  {author} {\bibinfo {author} {\bibfnamefont {T.}~\bibnamefont
  {Thiemann}},\ }\bibfield  {title} {\bibinfo {title} {{Quantum spin dynamics
  (qsd). 2.}},\ }\href {https://doi.org/10.1088/0264-9381/15/4/012} {\bibfield
  {journal} {\bibinfo  {journal} {Class. Quant. Grav.}\ }\textbf {\bibinfo
  {volume} {15}},\ \bibinfo {pages} {875} (\bibinfo {year}
  {1998}{\natexlab{b}})},\ \Eprint {https://arxiv.org/abs/gr-qc/9606090}
  {arXiv:gr-qc/9606090} \BibitemShut {NoStop}%
\bibitem [{\citenamefont {Giesel}\ and\ \citenamefont
  {Thiemann}(2007)}]{Giesel:2006uj}%
  \BibitemOpen
  \bibfield  {author} {\bibinfo {author} {\bibfnamefont {K.}~\bibnamefont
  {Giesel}}\ and\ \bibinfo {author} {\bibfnamefont {T.}~\bibnamefont
  {Thiemann}},\ }\bibfield  {title} {\bibinfo {title} {{Algebraic Quantum
  Gravity (AQG). I. Conceptual Setup}},\ }\href
  {https://doi.org/10.1088/0264-9381/24/10/003} {\bibfield  {journal} {\bibinfo
   {journal} {Class. Quant. Grav.}\ }\textbf {\bibinfo {volume} {24}},\
  \bibinfo {pages} {2465} (\bibinfo {year} {2007})},\ \Eprint
  {https://arxiv.org/abs/gr-qc/0607099} {arXiv:gr-qc/0607099} \BibitemShut
  {NoStop}%
\bibitem [{\citenamefont {Yang}\ \emph {et~al.}(2009)\citenamefont {Yang},
  \citenamefont {Ding},\ and\ \citenamefont {Ma}}]{Yang:2009fp}%
  \BibitemOpen
  \bibfield  {author} {\bibinfo {author} {\bibfnamefont {J.}~\bibnamefont
  {Yang}}, \bibinfo {author} {\bibfnamefont {Y.}~\bibnamefont {Ding}},\ and\
  \bibinfo {author} {\bibfnamefont {Y.}~\bibnamefont {Ma}},\ }\bibfield
  {title} {\bibinfo {title} {{\em Alternative quantization of the Hamiltonian
  in loop quantum cosmology II: Including the Lorentz term}},\ }\href
  {https://doi.org/10.1016/j.physletb.2009.10.072} {\bibfield  {journal}
  {\bibinfo  {journal} {Phys. Lett. B}\ }\textbf {\bibinfo {volume} {682}},\
  \bibinfo {pages} {1} (\bibinfo {year} {2009})},\ \Eprint
  {https://arxiv.org/abs/0904.4379} {arXiv:0904.4379 [gr-qc]} \BibitemShut
  {NoStop}%
\bibitem [{\citenamefont {Li}\ \emph {et~al.}(2018{\natexlab{a}})\citenamefont
  {Li}, \citenamefont {Singh},\ and\ \citenamefont {Wang}}]{Li:2018opr}%
  \BibitemOpen
  \bibfield  {author} {\bibinfo {author} {\bibfnamefont {B.-F.}\ \bibnamefont
  {Li}}, \bibinfo {author} {\bibfnamefont {P.}~\bibnamefont {Singh}},\ and\
  \bibinfo {author} {\bibfnamefont {A.}~\bibnamefont {Wang}},\ }\bibfield
  {title} {\bibinfo {title} {{Towards Cosmological Dynamics from Loop Quantum
  Gravity}},\ }\href {https://doi.org/10.1103/PhysRevD.97.084029} {\bibfield
  {journal} {\bibinfo  {journal} {Phys. Rev. D}\ }\textbf {\bibinfo {volume}
  {97}},\ \bibinfo {pages} {084029} (\bibinfo {year} {2018}{\natexlab{a}})},\
  \Eprint {https://arxiv.org/abs/1801.07313} {arXiv:1801.07313 [gr-qc]}
  \BibitemShut {NoStop}%
\bibitem [{\citenamefont {Li}\ \emph {et~al.}(2018{\natexlab{b}})\citenamefont
  {Li}, \citenamefont {Singh},\ and\ \citenamefont {Wang}}]{Li:2018fco}%
  \BibitemOpen
  \bibfield  {author} {\bibinfo {author} {\bibfnamefont {B.-F.}\ \bibnamefont
  {Li}}, \bibinfo {author} {\bibfnamefont {P.}~\bibnamefont {Singh}},\ and\
  \bibinfo {author} {\bibfnamefont {A.}~\bibnamefont {Wang}},\ }\bibfield
  {title} {\bibinfo {title} {{Qualitative dynamics and inflationary attractors
  in loop cosmology}},\ }\href {https://doi.org/10.1103/PhysRevD.98.066016}
  {\bibfield  {journal} {\bibinfo  {journal} {Phys. Rev. D}\ }\textbf {\bibinfo
  {volume} {98}},\ \bibinfo {pages} {066016} (\bibinfo {year}
  {2018}{\natexlab{b}})},\ \Eprint {https://arxiv.org/abs/1807.05236}
  {arXiv:1807.05236 [gr-qc]} \BibitemShut {NoStop}%
\bibitem [{\citenamefont {Li}\ \emph {et~al.}(2019)\citenamefont {Li},
  \citenamefont {Singh},\ and\ \citenamefont {Wang}}]{Li:2019ipm}%
  \BibitemOpen
  \bibfield  {author} {\bibinfo {author} {\bibfnamefont {B.-F.}\ \bibnamefont
  {Li}}, \bibinfo {author} {\bibfnamefont {P.}~\bibnamefont {Singh}},\ and\
  \bibinfo {author} {\bibfnamefont {A.}~\bibnamefont {Wang}},\ }\bibfield
  {title} {\bibinfo {title} {{Genericness of pre-inflationary dynamics and
  probability of the desired slow-roll inflation in modified loop quantum
  cosmologies}},\ }\href {https://doi.org/10.1103/PhysRevD.100.063513}
  {\bibfield  {journal} {\bibinfo  {journal} {Phys. Rev. D}\ }\textbf {\bibinfo
  {volume} {100}},\ \bibinfo {pages} {063513} (\bibinfo {year} {2019})},\
  \Eprint {https://arxiv.org/abs/1906.01001} {arXiv:1906.01001 [gr-qc]}
  \BibitemShut {NoStop}%
\bibitem [{\citenamefont {Li}\ \emph {et~al.}(2021)\citenamefont {Li},
  \citenamefont {Singh},\ and\ \citenamefont {Wang}}]{Li:2021mop}%
  \BibitemOpen
  \bibfield  {author} {\bibinfo {author} {\bibfnamefont {B.-F.}\ \bibnamefont
  {Li}}, \bibinfo {author} {\bibfnamefont {P.}~\bibnamefont {Singh}},\ and\
  \bibinfo {author} {\bibfnamefont {A.}~\bibnamefont {Wang}},\ }\bibfield
  {title} {\bibinfo {title} {{Phenomenological implications of modified loop
  cosmologies: an overview}},\ }\href
  {https://doi.org/10.3389/fspas.2021.701417} {\bibfield  {journal} {\bibinfo
  {journal} {Front. Astron. Space Sci.}\ }\textbf {\bibinfo {volume} {8}},\
  \bibinfo {pages} {701417} (\bibinfo {year} {2021})},\ \Eprint
  {https://arxiv.org/abs/2105.14067} {arXiv:2105.14067 [gr-qc]} \BibitemShut
  {NoStop}%
\bibitem [{\citenamefont {Li}\ \emph {et~al.}(2020{\natexlab{a}})\citenamefont
  {Li}, \citenamefont {Singh},\ and\ \citenamefont {Wang}}]{Li:2019qzr}%
  \BibitemOpen
  \bibfield  {author} {\bibinfo {author} {\bibfnamefont {B.-F.}\ \bibnamefont
  {Li}}, \bibinfo {author} {\bibfnamefont {P.}~\bibnamefont {Singh}},\ and\
  \bibinfo {author} {\bibfnamefont {A.}~\bibnamefont {Wang}},\ }\bibfield
  {title} {\bibinfo {title} {{Primordial power spectrum from the dressed metric
  approach in loop cosmologies}},\ }\href
  {https://doi.org/10.1103/PhysRevD.101.086004} {\bibfield  {journal} {\bibinfo
   {journal} {Phys. Rev. D}\ }\textbf {\bibinfo {volume} {101}},\ \bibinfo
  {pages} {086004} (\bibinfo {year} {2020}{\natexlab{a}})},\ \Eprint
  {https://arxiv.org/abs/1912.08225} {arXiv:1912.08225 [gr-qc]} \BibitemShut
  {NoStop}%
\bibitem [{\citenamefont {Li}\ \emph {et~al.}(2020{\natexlab{b}})\citenamefont
  {Li}, \citenamefont {Olmedo}, \citenamefont {Singh},\ and\ \citenamefont
  {Wang}}]{Li:2020mfi}%
  \BibitemOpen
  \bibfield  {author} {\bibinfo {author} {\bibfnamefont {B.-F.}\ \bibnamefont
  {Li}}, \bibinfo {author} {\bibfnamefont {J.}~\bibnamefont {Olmedo}}, \bibinfo
  {author} {\bibfnamefont {P.}~\bibnamefont {Singh}},\ and\ \bibinfo {author}
  {\bibfnamefont {A.}~\bibnamefont {Wang}},\ }\bibfield  {title} {\bibinfo
  {title} {{Primordial scalar power spectrum from the hybrid approach in loop
  cosmologies}},\ }\href {https://doi.org/10.1103/PhysRevD.102.126025}
  {\bibfield  {journal} {\bibinfo  {journal} {Phys. Rev. D}\ }\textbf {\bibinfo
  {volume} {102}},\ \bibinfo {pages} {126025} (\bibinfo {year}
  {2020}{\natexlab{b}})},\ \Eprint {https://arxiv.org/abs/2008.09135}
  {arXiv:2008.09135 [gr-qc]} \BibitemShut {NoStop}%
\bibitem [{\citenamefont {Bojowald}(2020)}]{Bojowald:2020nwa}%
  \BibitemOpen
  \bibfield  {author} {\bibinfo {author} {\bibfnamefont {M.}~\bibnamefont
  {Bojowald}},\ }\href {https://doi.org/10.1088/2514-3433/ab9c98} {\emph
  {\bibinfo {title} {{Foundations of Quantum Cosmology}}}}\ (\bibinfo
  {publisher} {IOP},\ \bibinfo {year} {2020})\BibitemShut {NoStop}%
\bibitem [{\citenamefont {Bojowald}(2011)}]{Bojowald:2011zzb}%
  \BibitemOpen
  \bibfield  {author} {\bibinfo {author} {\bibfnamefont {M.}~\bibnamefont
  {Bojowald}},\ }\href {https://doi.org/10.1007/978-1-4419-8276-6} {\emph
  {\bibinfo {title} {{Quantum cosmology}: {A Fundamental Description of the
  Universe}}}},\ Vol.\ \bibinfo {volume} {835}\ (\bibinfo {year}
  {2011})\BibitemShut {NoStop}%
\bibitem [{\citenamefont {Calcagni}(2017)}]{Calcagni:2017sdq}%
  \BibitemOpen
  \bibfield  {author} {\bibinfo {author} {\bibfnamefont {G.}~\bibnamefont
  {Calcagni}},\ }\href {https://doi.org/10.1007/978-3-319-41127-9} {\emph
  {\bibinfo {title} {{Classical and Quantum Cosmology}}}},\ Graduate Texts in
  Physics\ (\bibinfo  {publisher} {Springer},\ \bibinfo {year}
  {2017})\BibitemShut {NoStop}%
\bibitem [{\citenamefont {Jalalzadeh}\ and\ \citenamefont
  {Vargas~Moniz}(2022)}]{Jalalzadeh:2020bqu}%
  \BibitemOpen
  \bibfield  {author} {\bibinfo {author} {\bibfnamefont {S.}~\bibnamefont
  {Jalalzadeh}}\ and\ \bibinfo {author} {\bibfnamefont {P.}~\bibnamefont
  {Vargas~Moniz}},\ }\href {https://doi.org/10.1142/8540} {\emph {\bibinfo
  {title} {{Challenging Routes in Quantum Cosmology}}}}\ (\bibinfo  {publisher}
  {World Scientific},\ \bibinfo {year} {2022})\BibitemShut {NoStop}%
\bibitem [{\citenamefont {Ryan}(1972)}]{ryder:1972}%
  \BibitemOpen
  \bibfield  {author} {\bibinfo {author} {\bibfnamefont {M.}~\bibnamefont
  {Ryan}},\ }\href@noop {} {\emph {\bibinfo {title} {{Hamiltonian
  Cosmology}}}}\ (\bibinfo  {publisher} {Springer Berlin, Heidelberg},\
  \bibinfo {year} {1972})\BibitemShut {NoStop}%
\bibitem [{\citenamefont {Hartle}\ and\ \citenamefont
  {Hawking}(1983)}]{Hartle:1983ai}%
  \BibitemOpen
  \bibfield  {author} {\bibinfo {author} {\bibfnamefont {J.~B.}\ \bibnamefont
  {Hartle}}\ and\ \bibinfo {author} {\bibfnamefont {S.~W.}\ \bibnamefont
  {Hawking}},\ }\bibfield  {title} {\bibinfo {title} {{Wave Function of the
  Universe}},\ }\href {https://doi.org/10.1103/PhysRevD.28.2960} {\bibfield
  {journal} {\bibinfo  {journal} {Phys. Rev. D}\ }\textbf {\bibinfo {volume}
  {28}},\ \bibinfo {pages} {2960} (\bibinfo {year} {1983})}\BibitemShut
  {NoStop}%
\bibitem [{\citenamefont {Bayen}\ \emph {et~al.}(1977)\citenamefont {Bayen},
  \citenamefont {Flato}, \citenamefont {Fronsdal}, \citenamefont
  {Lichnerowicz},\ and\ \citenamefont {Sternheimer}}]{Bayen:1977pr}%
  \BibitemOpen
  \bibfield  {author} {\bibinfo {author} {\bibfnamefont {F.}~\bibnamefont
  {Bayen}}, \bibinfo {author} {\bibfnamefont {M.}~\bibnamefont {Flato}},
  \bibinfo {author} {\bibfnamefont {C.}~\bibnamefont {Fronsdal}}, \bibinfo
  {author} {\bibfnamefont {A.}~\bibnamefont {Lichnerowicz}},\ and\ \bibinfo
  {author} {\bibfnamefont {D.}~\bibnamefont {Sternheimer}},\ }\bibfield
  {title} {\bibinfo {title} {{Quantum Mechanics as a Deformation of Classical
  Mechanics}},\ }\href {https://doi.org/10.1007/BF00399745} {\bibfield
  {journal} {\bibinfo  {journal} {Lett. Math. Phys.}\ }\textbf {\bibinfo
  {volume} {1}},\ \bibinfo {pages} {521} (\bibinfo {year} {1977})}\BibitemShut
  {NoStop}%
\bibitem [{\citenamefont {Bayen}\ \emph
  {et~al.}(1978{\natexlab{a}})\citenamefont {Bayen}, \citenamefont {Flato},
  \citenamefont {Fronsdal}, \citenamefont {Lichnerowicz},\ and\ \citenamefont
  {Sternheimer}}]{Bayen:1977ha}%
  \BibitemOpen
  \bibfield  {author} {\bibinfo {author} {\bibfnamefont {F.}~\bibnamefont
  {Bayen}}, \bibinfo {author} {\bibfnamefont {M.}~\bibnamefont {Flato}},
  \bibinfo {author} {\bibfnamefont {C.}~\bibnamefont {Fronsdal}}, \bibinfo
  {author} {\bibfnamefont {A.}~\bibnamefont {Lichnerowicz}},\ and\ \bibinfo
  {author} {\bibfnamefont {D.}~\bibnamefont {Sternheimer}},\ }\bibfield
  {title} {\bibinfo {title} {{Deformation Theory and Quantization. 1.
  Deformations of Symplectic Structures}},\ }\href
  {https://doi.org/10.1016/0003-4916(78)90224-5} {\bibfield  {journal}
  {\bibinfo  {journal} {Annals Phys.}\ }\textbf {\bibinfo {volume} {111}},\
  \bibinfo {pages} {61} (\bibinfo {year} {1978}{\natexlab{a}})}\BibitemShut
  {NoStop}%
\bibitem [{\citenamefont {Bayen}\ \emph
  {et~al.}(1978{\natexlab{b}})\citenamefont {Bayen}, \citenamefont {Flato},
  \citenamefont {Fronsdal}, \citenamefont {Lichnerowicz},\ and\ \citenamefont
  {Sternheimer}}]{Bayen:1977hb}%
  \BibitemOpen
  \bibfield  {author} {\bibinfo {author} {\bibfnamefont {F.}~\bibnamefont
  {Bayen}}, \bibinfo {author} {\bibfnamefont {M.}~\bibnamefont {Flato}},
  \bibinfo {author} {\bibfnamefont {C.}~\bibnamefont {Fronsdal}}, \bibinfo
  {author} {\bibfnamefont {A.}~\bibnamefont {Lichnerowicz}},\ and\ \bibinfo
  {author} {\bibfnamefont {D.}~\bibnamefont {Sternheimer}},\ }\bibfield
  {title} {\bibinfo {title} {{Deformation Theory and Quantization. 2. Physical
  Applications}},\ }\href {https://doi.org/10.1016/0003-4916(78)90225-7}
  {\bibfield  {journal} {\bibinfo  {journal} {Annals Phys.}\ }\textbf {\bibinfo
  {volume} {111}},\ \bibinfo {pages} {111} (\bibinfo {year}
  {1978}{\natexlab{b}})}\BibitemShut {NoStop}%
\bibitem [{\citenamefont {Cordero}\ \emph {et~al.}(2011)\citenamefont
  {Cordero}, \citenamefont {Garcia-Compean},\ and\ \citenamefont
  {Turrubiates}}]{Cordero:2011xa}%
  \BibitemOpen
  \bibfield  {author} {\bibinfo {author} {\bibfnamefont {R.}~\bibnamefont
  {Cordero}}, \bibinfo {author} {\bibfnamefont {H.}~\bibnamefont
  {Garcia-Compean}},\ and\ \bibinfo {author} {\bibfnamefont {F.~J.}\
  \bibnamefont {Turrubiates}},\ }\bibfield  {title} {\bibinfo {title}
  {{Deformation quantization of cosmological models}},\ }\href
  {https://doi.org/10.1103/PhysRevD.83.125030} {\bibfield  {journal} {\bibinfo
  {journal} {Phys. Rev. D}\ }\textbf {\bibinfo {volume} {83}},\ \bibinfo
  {pages} {125030} (\bibinfo {year} {2011})},\ \Eprint
  {https://arxiv.org/abs/1102.4379} {arXiv:1102.4379 [hep-th]} \BibitemShut
  {NoStop}%
\bibitem [{\citenamefont {Vakili}\ \emph {et~al.}(2010)\citenamefont {Vakili},
  \citenamefont {Islamic Azad~U.}, \citenamefont {Pedram},\ and\ \citenamefont
  {Jalalzadeh}}]{Vakili:2010qf}%
  \BibitemOpen
  \bibfield  {author} {\bibinfo {author} {\bibfnamefont {B.}~\bibnamefont
  {Vakili}}, \bibinfo {author} {\bibfnamefont {C.}~\bibnamefont {Islamic
  Azad~U.}}, \bibinfo {author} {\bibfnamefont {P.}~\bibnamefont {Pedram}},\
  and\ \bibinfo {author} {\bibfnamefont {S.}~\bibnamefont {Jalalzadeh}},\
  }\bibfield  {title} {\bibinfo {title} {{Late time acceleration in a deformed
  phase space model of dilaton cosmology}},\ }\href
  {https://doi.org/10.1016/j.physletb.2010.03.016} {\bibfield  {journal}
  {\bibinfo  {journal} {Phys. Lett. B}\ }\textbf {\bibinfo {volume} {687}},\
  \bibinfo {pages} {119} (\bibinfo {year} {2010})},\ \Eprint
  {https://arxiv.org/abs/1003.1194} {arXiv:1003.1194 [gr-qc]} \BibitemShut
  {NoStop}%
\bibitem [{\citenamefont {Malekolkalami}\ \emph {et~al.}(2014)\citenamefont
  {Malekolkalami}, \citenamefont {Atazadeh},\ and\ \citenamefont
  {Vakili}}]{Malekolkalami:2014dca}%
  \BibitemOpen
  \bibfield  {author} {\bibinfo {author} {\bibfnamefont {B.}~\bibnamefont
  {Malekolkalami}}, \bibinfo {author} {\bibfnamefont {K.}~\bibnamefont
  {Atazadeh}},\ and\ \bibinfo {author} {\bibfnamefont {B.}~\bibnamefont
  {Vakili}},\ }\bibfield  {title} {\bibinfo {title} {{Late time acceleration in
  a non-commutative model of modified cosmology}},\ }\href
  {https://doi.org/10.1016/j.physletb.2014.11.003} {\bibfield  {journal}
  {\bibinfo  {journal} {Phys. Lett. B}\ }\textbf {\bibinfo {volume} {739}},\
  \bibinfo {pages} {400} (\bibinfo {year} {2014})},\ \Eprint
  {https://arxiv.org/abs/1411.3623} {arXiv:1411.3623 [gr-qc]} \BibitemShut
  {NoStop}%
\bibitem [{\citenamefont {Rashki}\ and\ \citenamefont
  {Jalalzadeh}(2015)}]{Rashki:2014noa}%
  \BibitemOpen
  \bibfield  {author} {\bibinfo {author} {\bibfnamefont {M.}~\bibnamefont
  {Rashki}}\ and\ \bibinfo {author} {\bibfnamefont {S.}~\bibnamefont
  {Jalalzadeh}},\ }\bibfield  {title} {\bibinfo {title} {{Holography from
  quantum cosmology}},\ }\href {https://doi.org/10.1103/PhysRevD.91.023501}
  {\bibfield  {journal} {\bibinfo  {journal} {Phys. Rev. D}\ }\textbf {\bibinfo
  {volume} {91}},\ \bibinfo {pages} {023501} (\bibinfo {year} {2015})},\
  \Eprint {https://arxiv.org/abs/1412.3950} {arXiv:1412.3950 [gr-qc]}
  \BibitemShut {NoStop}%
\bibitem [{\citenamefont {Jalalzadeh}\ \emph {et~al.}(2017)\citenamefont
  {Jalalzadeh}, \citenamefont {Capistrano},\ and\ \citenamefont
  {Moniz}}]{Jalalzadeh:2017jdo}%
  \BibitemOpen
  \bibfield  {author} {\bibinfo {author} {\bibfnamefont {S.}~\bibnamefont
  {Jalalzadeh}}, \bibinfo {author} {\bibfnamefont {A.~J.~S.}\ \bibnamefont
  {Capistrano}},\ and\ \bibinfo {author} {\bibfnamefont {P.~V.}\ \bibnamefont
  {Moniz}},\ }\bibfield  {title} {\bibinfo {title} {{Quantum deformation of
  quantum cosmology: A framework to discuss the cosmological constant
  problem}},\ }\href {https://doi.org/10.1016/j.dark.2017.09.011} {\bibfield
  {journal} {\bibinfo  {journal} {Phys. Dark Univ.}\ }\textbf {\bibinfo
  {volume} {18}},\ \bibinfo {pages} {55} (\bibinfo {year} {2017})},\ \Eprint
  {https://arxiv.org/abs/1709.09923} {arXiv:1709.09923 [gr-qc]} \BibitemShut
  {NoStop}%
\bibitem [{\citenamefont {Bina}\ \emph {et~al.}(2010)\citenamefont {Bina},
  \citenamefont {Jalalzadeh},\ and\ \citenamefont {Moslehi}}]{Bina:2010ir}%
  \BibitemOpen
  \bibfield  {author} {\bibinfo {author} {\bibfnamefont {A.}~\bibnamefont
  {Bina}}, \bibinfo {author} {\bibfnamefont {S.}~\bibnamefont {Jalalzadeh}},\
  and\ \bibinfo {author} {\bibfnamefont {A.}~\bibnamefont {Moslehi}},\
  }\bibfield  {title} {\bibinfo {title} {{Quantum Black Hole in the Generalized
  Uncertainty Principle Framework}},\ }\href
  {https://doi.org/10.1103/PhysRevD.81.023528} {\bibfield  {journal} {\bibinfo
  {journal} {Phys. Rev. D}\ }\textbf {\bibinfo {volume} {81}},\ \bibinfo
  {pages} {023528} (\bibinfo {year} {2010})},\ \Eprint
  {https://arxiv.org/abs/1001.0861} {arXiv:1001.0861 [gr-qc]} \BibitemShut
  {NoStop}%
\bibitem [{\citenamefont {Gamboa}\ \emph {et~al.}(2001)\citenamefont {Gamboa},
  \citenamefont {Loewe},\ and\ \citenamefont {Rojas}}]{Gamboa:2000yq}%
  \BibitemOpen
  \bibfield  {author} {\bibinfo {author} {\bibfnamefont {J.}~\bibnamefont
  {Gamboa}}, \bibinfo {author} {\bibfnamefont {M.}~\bibnamefont {Loewe}},\ and\
  \bibinfo {author} {\bibfnamefont {J.~C.}\ \bibnamefont {Rojas}},\ }\bibfield
  {title} {\bibinfo {title} {{Noncommutative quantum mechanics}},\ }\href
  {https://doi.org/10.1103/PhysRevD.64.067901} {\bibfield  {journal} {\bibinfo
  {journal} {Phys. Rev. D}\ }\textbf {\bibinfo {volume} {64}},\ \bibinfo
  {pages} {067901} (\bibinfo {year} {2001})},\ \Eprint
  {https://arxiv.org/abs/hep-th/0010220} {arXiv:hep-th/0010220} \BibitemShut
  {NoStop}%
\bibitem [{\citenamefont {Chaichian}\ \emph {et~al.}(2001)\citenamefont
  {Chaichian}, \citenamefont {Sheikh-Jabbari},\ and\ \citenamefont
  {Tureanu}}]{Chaichian:2000si}%
  \BibitemOpen
  \bibfield  {author} {\bibinfo {author} {\bibfnamefont {M.}~\bibnamefont
  {Chaichian}}, \bibinfo {author} {\bibfnamefont {M.~M.}\ \bibnamefont
  {Sheikh-Jabbari}},\ and\ \bibinfo {author} {\bibfnamefont {A.}~\bibnamefont
  {Tureanu}},\ }\bibfield  {title} {\bibinfo {title} {{Hydrogen atom spectrum
  and the Lamb shift in noncommutative QED}},\ }\href
  {https://doi.org/10.1103/PhysRevLett.86.2716} {\bibfield  {journal} {\bibinfo
   {journal} {Phys. Rev. Lett.}\ }\textbf {\bibinfo {volume} {86}},\ \bibinfo
  {pages} {2716} (\bibinfo {year} {2001})},\ \Eprint
  {https://arxiv.org/abs/hep-th/0010175} {arXiv:hep-th/0010175} \BibitemShut
  {NoStop}%
\bibitem [{\citenamefont {P\'erez-Pay\'an}\ \emph {et~al.}(2013)\citenamefont
  {P\'erez-Pay\'an}, \citenamefont {Sabido},\ and\ \citenamefont
  {Yee-Romero}}]{Perez-Payan:2011cvf}%
  \BibitemOpen
  \bibfield  {author} {\bibinfo {author} {\bibfnamefont {S.}~\bibnamefont
  {P\'erez-Pay\'an}}, \bibinfo {author} {\bibfnamefont {M.}~\bibnamefont
  {Sabido}},\ and\ \bibinfo {author} {\bibfnamefont {C.}~\bibnamefont
  {Yee-Romero}},\ }\bibfield  {title} {\bibinfo {title} {{Effects of deformed
  phase space on scalar field cosmology}},\ }\href
  {https://doi.org/10.1103/PhysRevD.88.027503} {\bibfield  {journal} {\bibinfo
  {journal} {Phys. Rev. D}\ }\textbf {\bibinfo {volume} {88}},\ \bibinfo
  {pages} {027503} (\bibinfo {year} {2013})},\ \Eprint
  {https://arxiv.org/abs/1111.6136} {arXiv:1111.6136 [hep-th]} \BibitemShut
  {NoStop}%
\bibitem [{\citenamefont {Perez-Payan}\ \emph {et~al.}(2014)\citenamefont
  {Perez-Payan}, \citenamefont {Sabido}, \citenamefont {Mena},\ and\
  \citenamefont {Yee-Romero}}]{Perez-Payan:2014kea}%
  \BibitemOpen
  \bibfield  {author} {\bibinfo {author} {\bibfnamefont {S.}~\bibnamefont
  {Perez-Payan}}, \bibinfo {author} {\bibfnamefont {M.}~\bibnamefont {Sabido}},
  \bibinfo {author} {\bibfnamefont {E.}~\bibnamefont {Mena}},\ and\ \bibinfo
  {author} {\bibfnamefont {C.}~\bibnamefont {Yee-Romero}},\ }\bibfield  {title}
  {\bibinfo {title} {{Analysis of Scalar Field Cosmology with Phase Space
  Deformations}},\ }\href {https://doi.org/10.1155/2014/958137} {\bibfield
  {journal} {\bibinfo  {journal} {Adv. High Energy Phys.}\ }\textbf {\bibinfo
  {volume} {2014}},\ \bibinfo {pages} {958137} (\bibinfo {year}
  {2014})}\BibitemShut {NoStop}%
\bibitem [{\citenamefont {L\'opez}\ \emph {et~al.}(2018)\citenamefont
  {L\'opez}, \citenamefont {Sabido},\ and\ \citenamefont
  {Yee-Romero}}]{Lopez:2017xaz}%
  \BibitemOpen
  \bibfield  {author} {\bibinfo {author} {\bibfnamefont {J.~L.}\ \bibnamefont
  {L\'opez}}, \bibinfo {author} {\bibfnamefont {M.}~\bibnamefont {Sabido}},\
  and\ \bibinfo {author} {\bibfnamefont {C.}~\bibnamefont {Yee-Romero}},\
  }\bibfield  {title} {\bibinfo {title} {{Phase space deformations in phantom
  cosmology}},\ }\href {https://doi.org/10.1016/j.dark.2017.12.006} {\bibfield
  {journal} {\bibinfo  {journal} {Phys. Dark Univ.}\ }\textbf {\bibinfo
  {volume} {19}},\ \bibinfo {pages} {104} (\bibinfo {year} {2018})},\ \Eprint
  {https://arxiv.org/abs/1711.01111} {arXiv:1711.01111 [gr-qc]} \BibitemShut
  {NoStop}%
\bibitem [{\citenamefont {Baumann}(2011)}]{Baumann:2009ds}%
  \BibitemOpen
  \bibfield  {author} {\bibinfo {author} {\bibfnamefont {D.}~\bibnamefont
  {Baumann}},\ }\bibfield  {title} {\bibinfo {title} {{TASI Lecture on
  Inflation}},\ }in\ \href {https://doi.org/10.1142/9789814327183_0010} {\emph
  {\bibinfo {booktitle} {{Physics of the large and the small, TASI 09,
  proceedings of the Theoretical Advanced Study Institute in Elementary
  Particle Physics, Boulder, Colorado, USA, 1-26 June 2009}}}}\ (\bibinfo
  {year} {2011})\ pp.\ \bibinfo {pages} {523--686},\ \Eprint
  {https://arxiv.org/abs/0907.5424} {arXiv:0907.5424 [hep-th]} \BibitemShut
  {NoStop}%
%%CITATION = ARXIV:0907.5424;%%
\bibitem [{\citenamefont {Agullo}\ and\ \citenamefont
  {Morris}(2015)}]{Agullo:2015tca}%
  \BibitemOpen
  \bibfield  {author} {\bibinfo {author} {\bibfnamefont {I.}~\bibnamefont
  {Agullo}}\ and\ \bibinfo {author} {\bibfnamefont {N.~A.}\ \bibnamefont
  {Morris}},\ }\bibfield  {title} {\bibinfo {title} {{Detailed analysis of the
  predictions of loop quantum cosmology for the primordial power spectra}},\
  }\href {https://doi.org/10.1103/PhysRevD.92.124040} {\bibfield  {journal}
  {\bibinfo  {journal} {Phys. Rev. D}\ }\textbf {\bibinfo {volume} {92}},\
  \bibinfo {pages} {124040} (\bibinfo {year} {2015})},\ \Eprint
  {https://arxiv.org/abs/1509.05693} {arXiv:1509.05693 [gr-qc]} \BibitemShut
  {NoStop}%
\bibitem [{\citenamefont {Shahalam}\ \emph {et~al.}(2017)\citenamefont
  {Shahalam}, \citenamefont {Sharma}, \citenamefont {Wu},\ and\ \citenamefont
  {Wang}}]{Shahalam:2017wba}%
  \BibitemOpen
  \bibfield  {author} {\bibinfo {author} {\bibfnamefont {M.}~\bibnamefont
  {Shahalam}}, \bibinfo {author} {\bibfnamefont {M.}~\bibnamefont {Sharma}},
  \bibinfo {author} {\bibfnamefont {Q.}~\bibnamefont {Wu}},\ and\ \bibinfo
  {author} {\bibfnamefont {A.}~\bibnamefont {Wang}},\ }\bibfield  {title}
  {\bibinfo {title} {{Preinflationary dynamics in loop quantum cosmology:
  Power-law potentials}},\ }\href {https://doi.org/10.1103/PhysRevD.96.123533}
  {\bibfield  {journal} {\bibinfo  {journal} {Phys. Rev. D}\ }\textbf {\bibinfo
  {volume} {96}},\ \bibinfo {pages} {123533} (\bibinfo {year} {2017})},\
  \Eprint {https://arxiv.org/abs/1710.09845} {arXiv:1710.09845 [gr-qc]}
  \BibitemShut {NoStop}%
\bibitem [{\citenamefont {Xiao}(2020)}]{Xiao:2020olb}%
  \BibitemOpen
  \bibfield  {author} {\bibinfo {author} {\bibfnamefont {K.}~\bibnamefont
  {Xiao}},\ }\bibfield  {title} {\bibinfo {title} {{Tachyon field in loop
  cosmology}},\ }\href {https://doi.org/10.1016/j.physletb.2020.135859}
  {\bibfield  {journal} {\bibinfo  {journal} {Phys. Lett. B}\ }\textbf
  {\bibinfo {volume} {811}},\ \bibinfo {pages} {135859} (\bibinfo {year}
  {2020})}\BibitemShut {NoStop}%
\bibitem [{\citenamefont {Zhu}\ \emph {et~al.}(2017)\citenamefont {Zhu},
  \citenamefont {Wang}, \citenamefont {Cleaver}, \citenamefont {Kirsten},\ and\
  \citenamefont {Sheng}}]{Zhu:2017jew}%
  \BibitemOpen
  \bibfield  {author} {\bibinfo {author} {\bibfnamefont {T.}~\bibnamefont
  {Zhu}}, \bibinfo {author} {\bibfnamefont {A.}~\bibnamefont {Wang}}, \bibinfo
  {author} {\bibfnamefont {G.}~\bibnamefont {Cleaver}}, \bibinfo {author}
  {\bibfnamefont {K.}~\bibnamefont {Kirsten}},\ and\ \bibinfo {author}
  {\bibfnamefont {Q.}~\bibnamefont {Sheng}},\ }\bibfield  {title} {\bibinfo
  {title} {{Pre-inflationary universe in loop quantum cosmology}},\ }\href
  {https://doi.org/10.1103/PhysRevD.96.083520} {\bibfield  {journal} {\bibinfo
  {journal} {Phys. Rev. D}\ }\textbf {\bibinfo {volume} {96}},\ \bibinfo
  {pages} {083520} (\bibinfo {year} {2017})},\ \Eprint
  {https://arxiv.org/abs/1705.07544} {arXiv:1705.07544 [gr-qc]} \BibitemShut
  {NoStop}%
\end{thebibliography}%

\end{document}